\UseRawInputEncoding
\documentclass[twocolumn]{aastex631}

\newcommand{\Msu}{$M_{\odot}$ }

\newcommand{\Zsun}{$Z_{\odot}$}

\newcommand{\hi}{{\rm H\,}{{\sc i}}}
 
\newcommand{\his}{{\rm H\,}{{\sc i }}}

\accepted{\today}

\submitjournal{ApJ}

\graphicspath{{/astro/dust_kg/jduval/METAL/}{DEPLETIONS/}}
\begin{document}

\title{METAL: The Metal Evolution, Transport, and Abundance in the Large Magellanic Cloud Hubble program. III. Interstellar Depletions, Dust-to-Metal, and Dust-to-Gas Ratios Versus Metallicity}
\author[0000-0001-6326-7069]{Julia Roman-Duval}
\affiliation{Space Telescope Science Institute\\
3700 San Martin Drive \\
Baltimore, MD21218, USA}
\author[0000-0003-1892-4423]{Edward B. Jenkins}
\affiliation{Department of Astrophysical Sciences\\
Peyton Hall, Princeton University\\
Princeton, NJ 08544-1001 USA}
\author[0000-0003-0789-9939]{Kirill Tchernyshyov}
\affiliation{Department of Astronomy\\
Box 351580, University of Washington\\
Seattle, WA 98195, USA}
\author[0000-0001-7959-4902]{Christopher J.R. Clark}
\affiliation{Space Telescope Science Institute\\
3700 San Martin Drive \\
Baltimore, MD21218, USA}
\author[0000-0003-2082-1626]{Annalisa De Cia}
\affiliation{Department of Astronomy, University of Geneva\\
Chemin Pegasi 51\\
1290 Versoix, Switzerland }
\author[0000-0001-5340-6774]{Karl Gordon}
\affiliation{Space Telescope Science Institute\\
3700 San Martin Drive \\
Baltimore, MD21218, USA}
\author[0000-0002-4646-7509]{Aleksandra Hamanowicz}
\affiliation{Space Telescope Science Institute\\
3700 San Martin Drive \\
Baltimore, MD21218, USA}
\author[0000-0002-7716-6223]{Vianney Lebouteiller}
\affiliation{AIM, CEA, CNRS, UniversitŽ Paris-Saclay, UniversitŽ Paris Diderot, Sorbonne Paris CitŽ\\
F-91191 Gif-sur-Yvette, France}
\author[0000-0002-9946-4731]{Marc Rafelski}
\affiliation{Space Telescope Science Institute\\
3700 San Martin Drive \\
Baltimore, MD21218, USA}
\author[0000-0002-4378-8534]{Karin Sandstrom}
\affiliation{Center for Astrophysics and Space Sciences, Department of Physics\\
University of California\\
9500 Gilman Drive\\
La Jolla, San Diego, CA 92093, USA}
\author[0000-0002-0355-0134]{Jessica Werk}
\affiliation{Department of Astronomy\\
Box 351580, University of Washington\\
Seattle, WA 98195, USA}
\author[0000-0002-9912-6046]{Petia Yanchulova Merica-Jones}
\affiliation{Space Telescope Science Institute\\
3700 San Martin Drive \\
Baltimore, MD21218, USA}

\begin{abstract}
The metallicity and gas density dependence of interstellar depletions, the dust-to-gas (D/G), and dust-to-metal (D/M) ratios have important implications for how accurately we can trace the chemical enrichment of the universe; either by using FIR dust emission as a tracer of the ISM; or by using spectroscopy of damped Lyman-$\alpha$ systems (DLAs) to measure chemical abundances over a wide range of redshifts. We collect and compare large samples of depletion measurements in the Milky Way (MW), LMC (Z=0.5\Zsun), and SMC (Z=0.2\Zsun). The relation between the depletions of different elements do not strongly vary between the three galaxies, implying that abundance ratios should trace depletions accurately down to 20\% solar metallicity. From the depletions, we derive D/G and D/M. The D/G increases with density, consistent with the more efficient accretion of gas-phase metals onto dust grains in the denser ISM. For $\log$ N(H) $>$ 21 cm$^{-2}$, the depletion of metallicity tracers (S, Zn) exceeds $-$0.5 dex, even at 20\% solar metallicity. The gas fraction of metals increases from the MW to the LMC (factor 3) and SMC (factor 6), compensating the reduction in total heavy element abundances and resulting in those three galaxies having the same neutral gas-phase metallicities. The D/G derived from depletions are a factor of 2 (LMC) and 5 (SMC) higher than the D/G derived from FIR, 21 cm, and CO emission, likely due to the combined uncertainties on the dust FIR opacity and on the depletion of carbon and oxygen. 
\end{abstract}



\section{Introduction}\label{introduction}

\indent Over a galaxy's lifetime, metals are produced in stars and deposited into the interstellar medium (ISM). These metals cycle between different phases of the ISM: some remain in the gas at different temperatures and pressures, others are locked into dust, and others are ejected through galactic winds into the circumgalactic medium (CGM), where they can rain back down into the ISM \citep{oppenheimer2008}. This incessant cycle of material between stars, interstellar gas and dust, and galaxy halos drives galaxy evolution. A critical, yet poorly understood, aspect of this baryon cycle is the depletion of metals from the gas to the dust phase via dust formation, and vice versa, the return of heavy elements from the dust to the gas phase via dust destruction. The parameters describing the lifecycle of metals in the neutral ISM are the dust-to-metal mass ratio (D/M, the mass fraction of metals locked up in dust grains) and the dust-to-gas mass ratio (D/G = D/M $\times$ Z, where Z is the metallicity).\\
\indent D/M and D/G are theoretically expected to vary with metallicity \citep[e.g., ][]{asano2013, feldmann2015, zhukovska2016, mattsson2020}. Above a critical metallicity at which the dust input rate from evolved stars (AGB + supernovae) and ISM dust growth balances the dust destruction by supernova (SN) shockwaves and dilution by inflows of pristine gas, the D/M is predicted to be high with most metals locked in the dust-phase. Below this critical metallicity, dust growth in the ISM is not efficient enough to counter balance destruction and dilution effects. In this case, the D/M is expected to be low and determined by the input of stellar dust sources.  Assuming the fiducial parameters in \citet{feldmann2015}, including a molecular depletion time of 2 Gyr \citep{bigiel2008} and timescale of dust growth in the ISM at solar metallicity of 500 Myr for volume densities of 100 cm$^{-3}$ \citep{hirashita2000, asano2013}, the critical metallicity separating low and high D/M and D/G is about 10\%---15\% solar. \\
\indent D/G and D/M can be measured in nearby galaxies using two distinct approaches and observational techniques. The gas and dust content of nearby galaxies can be estimated using emission-based tracers, specifically FIR emission to trace interstellar dust, and 21 cm and CO rotational emission to trace gas. Their ratio provides D/G, and if the metallicity of the system is known, also D/M. Alternatively, D/M and D/G can be estimated from depletion measurements using UV spectroscopy of interstellar absorption lines. The depletion of element X, $\delta$(X), is the logarithm of the fraction of X in the gas-phase, and is given by:
 
\begin{equation}\label{dep_equation}
\delta(X)  = \log_{10} \left ( \frac{N(\mathrm{X})}{N(\mathrm{H})} \right) - \log_{10} \left (\frac{\mathrm{X}}{H} \right )_{\mathrm{tot}}
\end{equation}
 
\noindent where $(X/H)_{\mathrm{tot}}$ are total ISM abundances, assumed to equate the abundances in the photospheres of young stars recently formed out of the ISM. Knowing depletions, and therefore also the fraction of metals in the dust, for elements that are the main constituents of dust yields D/M and D/G.\\
\indent Variations in D/M and D/G with metallicity, and how well such variations can be observationally constrained, have important implications for galaxy evolution and how accurately we can track the chemical enrichment of the universe. First, a comprehensive understanding of the variations of D/M and D/G with metallicity is required to estimate gas masses based on far-infrared dust emission in both nearby \citep{bolatto2011, schruba2011} and distant \citep{rowlands2014} galaxies. Second, understanding depletion patterns is critical to the interpretation of gas-phase abundance measurements in damped Lyman-$\alpha$ systems (DLAs). DLAs are neutral gas absorption systems with $\log$ N(H) $>$ 20.3 cm$^{-2}$ observed over a wide range of redshifts using quasar absorption spectroscopy \citep[e.g., ][]{rafelski2012, quiret2016, decia2018a}. Thanks to their \his and metallic absorption lines, DLAs trace the chemical enrichment of the universe over cosmic times, and carry the majority of metals at high redshift \citep{peroux2020}. However, gas-phase abundance measurements in DLAs have to be corrected for the depletion of metals from the gas to the dust phase, particularly at metallicities $>$1\% solar, and thus tracking the chemical enrichment of the universe through DLA spectroscopy requires understanding how the fraction of metals in the dust-phase varies with metallicity. This can be understood in nearby galaxies such as the Milky Way and Magellanic Clouds, where gas-phase, stellar, and ionized gas abundances can be measured.\\
 \indent On the one hand, measurements of D/M and D/G in nearby galaxies as a function of metallicity from observations using FIR, HI 21 cm and CO emission to trace dust, atomic, and molecular gas \citep[e.g.,][]{herrera-camus2012, remy-ruyer2014, devis2019} confirm the theoretical prediction that the D/G sharply decreases at a metallicity of 10\%---20\% solar \citep[see Figure 9a in the review by ][and references therein]{galliano2018}. The emission-based D/G measurements in nearby galaxies follow the model tracks from \citet{feldmann2015}, with the best agreement obtained for the parameter $\gamma$ $=$ 3$\times10^4$, where $\gamma$ is the ratio of the molecular gas consumption time by star-formation to the timescale for dust growth in the ISM. On the other hand, Figure 9 in \citet{galliano2018} shows that D/G measurements obtained from abundance ratios (in particular [Zn/Fe], see \citealt{decia2016}) in DLAs over a wide range of redshifts follow a different trend, close to linear with metallicity.  \\
\indent This tension between measurements of D/G obtained with emission-based tracers (FIR, 21 cm, CO) in nearby galaxies, rest-frame UV absorption in DLAs, and chemical evolution models could be explained by several factors. First, emission-based tracers suffer from substantial degeneracies and systematics: dust mass estimates are degenerate with the assumed FIR opacity of dust, which has been shown to vary \citep{stepnik2003, kohler2012, demyk2017} and is not well constrained observationally \citep{clark2019}. In estimating dust masses from the FIR, there is also a potential bias (under-estimation of the dust mass) due to the integrated nature of the measurement of dust surface densities that can vary on small scales \citep{galliano2011}. Gas masses estimated from 21 cm and CO emission also suffer from substantial systematics. The molecular gas mass estimates rely on an assumed CO-to-H$_2$ conversion factor \citep{bolatto2013}, which is also poorly constrained and degenerate with D/G measurements \citep{RD2014}. Another potential issue in estimating atomic gas masses from 21 cm emission is that masses are often estimated from integrated measurements associated with a region that is spatially more extended than the region detected in dust emission (either on the sky or along the line of sight) , leading to a possible over-estimation of the gas mass. Thus, the systematic uncertainty on D/G estimates based on emission tracers could very well amount to a factor of several, perhaps up to an order of magnitude, and the effects describe above would preferentially under-estimate the D/G. \\
\indent Second, the relation between depletions and abundance ratios in DLAs is calibrated on depletion measurements obtained in the Milky Way at solar metallicity \citep[a local calibration is required for at least one element, usually Zn, see][]{decia2016}. It is possible that nucleosynthetic effects modify this relation at low metallicity. Indeed, Zn could behave like an $\alpha$-process element \citep{ernandes2018}, with the stellar [Zn/Fe] ratio being enhanced in some stellar populations in an age and metallicity-dependent way \citep{duffau2017, dasilveira2018, delgado-mena2019}. Based on a small sample in the Large Magellanic Cloud \citep[LMC, 50\% solar metallicity,][]{russell1992} and the Small Magellanic Cloud \citep[SMC, 20\% solar metallicity,][]{russell1992}, \citet{decia2018b} show that the calibration of iron depletions, which correlate tightly with the depletions of other elements \citep{jenkins2009, RD2021}, as a function of [Zn/Fe] does not appear to change significantly between the Milky Way, LMC, and SMC, where interstellar depletions can be estimated from the gas and stellar abundances \citep{jenkins2009, tchernyshyov2015, decia2018b, RD2021}, as opposed to inferred from abundance ratios.  Nevertheless, only a few depletion measurements were available in the LMC until the METAL (GO-14675) large HST program obtained UV spectra of 32 sight-lines in the LMC \citep{RD2019, RD2021}.   \\
\indent In this paper, we compile recent depletion measurements in the Milky Way, LMC, and SMC and compare the relations between depletions of different elements and their abundance ratios between these three galaxies. From the depletions, we compute D/G and D/M and examine the relation between depletions, D/M, D/G, and hydrogen column density, which has been shown to be a driver of the D/M and D/G \citep{RD2017, chiang2018, chiang2021, RD2021}. Additionally, we examine the metallicity dependence of depletions, D/M, and D/G by also including D/G estimates in nearby galaxies obtained from FIR measurements.The results presented in this paper lay the groundwork for deriving calibrations of depletions as a function of abundance ratios that can be applied to DLAs in order to estimate the metal and dust content of the universe over cosmic times, which will be presented in the upcoming METAL IV paper (Roman-Duval et al., in prep).\\
\indent The paper is organized as follows. In Section \ref{compilation_section}, we present the details of the depletion measurements compiled in this paper. The depletions of different elements are compared in Section \ref{deps_vs_fe}. The derivation of D/M and D/G is presented in Section \ref{computing_dh} and the relation between depletions, D/M, D/G and hydrogen column density is examined in Section \ref{section_nh}. We infer the dust composition from depletions in the MW, LMC, and SMC in Section \ref{composition_section}. We examine the metallicity dependence of D/G based on the new depletion measurements in Section \ref{section_dg_vs_Z}. Results are summarized in the conclusion (Section \ref{conclusion}).

\section{Interstellar depletions from the literature}\label{compilation_section}

\indent In order to perform a comparison of depletions and their environmental variations between the Milky Way (MW), LMC, and SMC, we compile gas-phase abundance and depletion measurements in those galaxies from the literature. Depletions for element X are computed from gas-phase column density measurements, $\log N(X)$, and column density of hydrogen N(H) = N(\hi) + 2N(H$_2$)  following Equation \ref{dep_equation}. \\
\indent Depletions for different elements are observed to tightly correlate with each other, as observed by \citet{jenkins2009}. They introduced the $F_*$ parameter to describe the collective advancement of depletions in the Milky Way, with $F_*$ $=$ 0 corresponding to the least depleted sight-lines in the MW with log N(H) $>$ 19.5 cm$^{-2}$ (implying negligible ionization corrections) and $F_*$ $=$ 1 corresponding to the most depleted velocity component toward $\zeta$ Oph. Following \citet{jenkins2009}, the depletion of element X can be modeled from $F_*$ by:

\begin{equation}\label{fstar_equation}
 \delta(\mathrm{X}) = A_{\mathrm{X}} (F_*-z_{\mathrm{X}}) + B_{\mathrm{X}}
 \end{equation}
 
 \noindent where the $A_{\mathrm{X}}$, $B_{\mathrm{X}}$ and $z_{\mathrm{X}}$ coefficients are obtained from examining and fitting the relation between depletion measurements for different elements toward a sufficiently large sample of sight-lines. The term $z_{\mathrm{X}}$ is introduced to remove the covariance between errors on the slope ($A_{\mathrm{X}}$) and intercept ($B_{\mathrm{X}}$) of the relation. The $F_*$ parameter is critical in inferring depletions for elements when they cannot be measured, and thus for estimating the dust-to-metal and dust-to-gas ratios in different systems.\\
\indent In the following sections, we present the compilation of depletion measurements, as well as estimate of $A_{\mathrm{X}}$, $B_{\mathrm{X}}$ and $z_{\mathrm{X}}$, in the MW, LMC, and SMC. \\

\subsection{Milky Way}

\indent In the Milky Way, \citet[][their Table 7]{jenkins2009} provide a comprehensive compilation of gas-phase abundances and interstellar depletions measured from Copernicus and HST spectra. Of the 276 sight-lines and velocity components included in the \citet{jenkins2009} study (their Table 2), we select the 226 objects with robustly determined \his column densities, and determined H$_2$ column densities or upper limits on N(H$_2$) consistent with H$_2$ fractions, defined as 2N(H$_2$)/(N(\hi) + 2N(H$_2$)), lower than 10\%. For each of those sight-lines and velocity components, we only retain gas-phase column density measurements ($\log N(X)$) and discard upper or lower limits on $\log N(X)$, where X is one of the 17 elements included in the \citet{jenkins2009} study (C, N, O, Mg, Si, P, S, Cl, Cr, Mn, Fe, Ni, Cu, Zn, Ge, Kr, Ti). We note that the remaining determinations sample the parameter space as well as the full sample (including limits). Additionally, we only include sight-lines with $\log$ N(H) $>$ 19.5 cm$^{-2}$, because sight-lines with lower hydrogen column densities are susceptible to substantial ionization effects, making their abundance measurements unreliable \citep{jenkins2009}.\\
\indent For Zn, \citet{jenkins2009} assumed the oscillator strengths from \citet{morton2003}. However, the more recent depletion studies in the LMC and SMC used the newer, preferred oscillator strength from \citet{kisielius2015}. To homogenize the measurements, we therefore corrected the Zn column densities and depletions reported in \citet{jenkins2009} by $-$0.1 dex, which is the difference between the oscillator strengths for the Zn II $\lambda \lambda$ 2026, 2062 transitions in \citet{morton2003} and \citet{kisielius2015}. \\
\indent Table 4 of \citet{jenkins2009} provides the $A_{\mathrm{X}}$, $B_{\mathrm{X}}$, and $z_{\mathrm{X}}$ coefficients describing the relation between $\delta$(X) and $F_*$ for the Milky Way. We updated the zero-point $B_{\mathrm{Zn}}$ of this relation for Zn according to the correction performed on the depletion measurements to account for the more recent oscillator strength for the Zn lines. The $A_{\mathrm{X}}$, $B_{\mathrm{X}}$, and $z_{\mathrm{X}}$ coefficients for the MW are summarized in Table 2.
\indent Lastly, the total abundances for the Milky Way are listed in \citet{jenkins2009} and summarized in Table 1. \\

\subsection{Large Magellanic Cloud}

\indent In the LMC, \citet{RD2021} recently obtained gas-phase abundances and depletions for a large sample of sight-lines observed with HST/STIS and COS as part of the METAL (GO-14675) large program \citep{RD2019}. The study includes most of the major constituents of dust and other heavy elements commonly used as metallicity tracers (Mg, Si, S, Cr, Fe, Ni, Cu, Zn, Ti), but not C and O, for which UV transitions are either too saturated or too weak to be detected in the LMC. With 32 sight-lines with measured depletions for most of the elements listed above toward each sight-line, this is the most comprehensive sample available in this galaxy, and the one we use in this analysis. As for the MW, we only retain detections (not limits). The column density, abundance, and depletion measurements, along with the sight-line hydrogen column density, are listed in Table 5 of \citet{RD2021}.\\
\indent  We note that \citet{tchernyshyov2015} measured depletions toward a common sample of LMC sight-lines with FUSE and COS/NUV. Since the spectral resolution of STIS, predominantly used in the \citet{RD2021} study, is superior to the spectral resolution of both COS/NUV and FUSE, and since the \citet{tchernyshyov2015} sample is included in the \citet{RD2021} sample, we do not make use of their results in this analysis. Nonetheless, \citet{RD2021} did perform a comparison of the depletions obtained by both studies and found them to be in general agreement, within errors. \\
\indent The $A_{\mathrm{X}}$, $B_{\mathrm{X}}$ and $z_{\mathrm{X}}$ coefficients for the LMC are not directly available in \citet{RD2021}. Indeed, since the $F_*$ scale is tied to the particular sight lines used to anchor the $F_*$ $=$ 0 and 1 extremes, it is not possible to use the same normalizations of the $F_*$ scale in galaxies other than the MW. Therefore, similar to the computation of $F_*$ in SMC by \citet{jenkins2017}, the $F_*$ parameter in the LMC is given by:

\begin{equation}
F_* = \frac{\delta(\mathrm{Fe}) - B_{\mathrm{Fe}}}{A_{\mathrm{Fe}}} + z_{\mathrm{Fe}}
\label{fstar_definition}
\end{equation}

\noindent where $A_{\mathrm{Fe}}$ = $-$1.285 , $B_{\mathrm{Fe}}$ = $-$1.513 , and $z_{\mathrm{Fe}}$= 0.437 are the coefficients of the linear relation between $\delta$(Fe) and $F_*$ in the
Milky Way given in Table 4 of \citet{jenkins2009}. Fe was chosen to tie the $F_*$ scale in different galaxies since Fe depletions can generally be derived easily for all sight-lines. Then, we combine Equation \ref{fstar_definition} and the relation between $\delta$(Fe) and $\delta$(X), of the form $\delta$(X) $=$ $a_{\mathrm{X}}$($\delta$(Fe) - $\zeta_{\mathrm{X}}$) + $b_{\mathrm{X}}$,  given in Table 7 of \citet{RD2021} to compute $A_{\mathrm{X}}$, $B_{\mathrm{X}}$ and $z_{\mathrm{X}}$. We note that $\zeta_{\mathrm{X}}$ correspond to $z_{\mathrm{X}}$ in Table 7 of \citet{RD2021}.

\begin{equation}
A_{\mathrm{X}} = -1.285 a_{\mathrm{X}}
\end{equation}

\begin{equation}
B_{\mathrm{X}} =  b_{\mathrm{X}}
\end{equation}

\begin{equation}
z_{\mathrm{X}} = -\frac{\zeta_{\mathrm{X}} + 0.951}{1.285}
\end{equation}

\noindent The $A_{\mathrm{X}}$, $B_{\mathrm{X}}$ and $z_{\mathrm{X}}$ coefficients for the LMC are given in Table 2. Total abundances assumed to derive depletions in the LMC are identical to those used in \citet{RD2021} and are listed in Table 1. In particular abundances for all elements except S, Ti, Cu were taken from \citet{tchernyshyov2015}, who used a statistical technique to combine disparate measurements of stellar abundances in the Magellanic Clouds. \citet{tchernyshyov2015}, and abundances for those elements were taken from different publications (see Table 1).

\subsection{Small Magellanic Cloud}
\indent Depletions for 9 elements (Mg, Si, S, Cr, Fe, Ni, Cu, Zn, Ti) were obtained by \citet{jenkins2017} toward 18 stars in the SMC observed with HST/STIS. We use these measurements in our analysis. The column density, abundance, and depletion measurements, along with sight-lines information such as $\log$ N(\hi), $\log$ N(H$_2$) are given in their Table 3. \\
\indent As for the LMC, \citet{tchernyshyov2015} obtained depletions for 13 stars in the SMC. Eight of those sight-lines were observed with COS/NUV, yielding depletions for Si, Zn, Cr, Fe, and P.  Of those 8 stars, 4 were re-observed with STIS by \citet{jenkins2017}, which resulted in more accurate measurements for Si, Zn, Cr, Fe and new measurements for other elements (e.g., Mg). The other 5 SMC sight-lines from \citet{tchernyshyov2015} were observed with FUSE, providing depletions for Fe and P only. Since our analysis relies on samples with the full set of depletions measurements, including the major constituents of dust such as Mg, Si, Fe, Ni, as well as commonly used metallicity and depletion tracers such as S and Zn, we do not include the depletion measurements from \citet{tchernyshyov2015}, which targeted a smaller subset of elements.\\
\indent \citet{jenkins2017} estimated $F_*$ following Equation \ref{fstar_definition}, with the resulting $F_*$ values for the SMC sight-lines listed in their Table 5. The $A_{\mathrm{X}}$, $B_{\mathrm{X}}$ and $z_{\mathrm{X}}$ for the SMC are given in their Table 6, and summarized in Table 2.\\
\indent When available, we assume the same total abundances as \citet{jenkins2017} for the SMC (their Table 2).  \citet{jenkins2017} did not include C and O, and for those elements we assume total abundances from \citet{tchernyshyov2015}. Total abundances for the SMC and associated references are summarized in Table 1.\\

\begin{deluxetable*}{cccccccc}
\tablenum{1}
\tablecaption{Reference abundances of young stars (a proxy for the ISM total abundances) in the MW, LMC, SMC}\label{tab:reference_abundances}
\tablewidth{0pt}
\tablehead{
\colhead{Element} & \colhead{W$_{\mathrm{X}}$}  & \colhead{MW 12+$\log$(X/H)$_{\mathrm{tot}}$} & \colhead{Ref\tablenotemark{a}} & \colhead{LMC 12+$\log$(X/H)$_{\mathrm{tot}}$} & \colhead{Ref\tablenotemark{a}} & \colhead{SMC 12+$\log$(X/H)$_{\mathrm{tot}}$} & \colhead{Ref\tablenotemark{a}}\\
}
%\decimalcolnumbers
\startdata
C   &    12.01 & 8.46  & 1  & 7.94  & 2  &     7.52 & 2  \\
O     &  16.0  & 8.76  & 1  & 8.50  & 2     &   8.14 & 2 \\                               
Mg    &  24.3  & 7.62 & 1  & 7.26  & 2    &     6.95   & 6 \\                   
Si    &  28.1  & 7.61  & 1 & 7.35  & 2   &     6.86   & 6      \\      
S    &   32.06 & 7.26 & 1 & 7.13  & 3   &    6.47 & 6  \\    
Ti    &  47.87 & 5.00 & 1  & 4.76  & 4    &  4.30 & 6  \\   
Cr    &  52.0  & 5.72 & 1  & 5.37 & 2   & 4.99 & 6 \\ 
Fe   &   55.85  &7.54 & 1  &  7.32   &2  &      6.85 & 6\\               
Ni   &   58.7  & 6.29 & 1  & 5.92  & 2   &     5.57  & 6   \\           
Cu   &   63.55 & 4.34 & 1 &  3.79  &  5  &    \nodata & \nodata  \\  
Zn   &   65.4  & 4.70 & 1  &  4.31  & 2 &  3.91 & 6 \\ 
\enddata
\tablenotetext{a}{(1) \citet{jenkins2009}, who adopt solar abundances from \citet{lodders2003}; (2) \citet{tchernyshyov2015}; (3) 12 + log(S/H) $=$ [S/Fe] + (S/Fe)$_{\odot}$ + 12 + log(Fe/H) with [S/Fe] from \citet{hill1995}, 12 + log(Fe/H) from (2), and (S/Fe)$_{\odot}$ from \citet{lodders2021} ; (4) \citet{welty2010}; (5) \citet{asplund2009} scaled by factor 0.5 ($-$0.3 dex); (6) \citet{jenkins2017}, who scale proto-solar abundances from \citet{lodders2003} by a factor 0.22 ($-$0.66 dex)}
\end{deluxetable*}

\begin{deluxetable*}{cccc|ccc|ccc}
\tablenum{2}
\tablecaption{$A_{\mathrm{X}}$, $B_{\mathrm{X}}$, and $z_{\mathrm{X}}$ coefficients relating depletions and $F_*$ in the MW, LMC, and SMC} \label{tab:fstar_coeffs}
\tablewidth{0pt}
\tablehead{
\colhead{Element} &  \multicolumn{3}{c}{$A_{\mathrm{X}}$} & \multicolumn{3}{c}{$B_{\mathrm{X}}$} & \multicolumn{3}{c}{$z_{\mathrm{X}}$}\\
\cline{2-4}  \cline{5-7} \cline{8-10}
& \colhead{MW} & \colhead{LMC} & \colhead{SMC} & \colhead{MW} & \colhead{LMC} & \colhead{SMC} & \colhead{MW} & \colhead{LMC} &\colhead{SMC} 
}
%\decimalcolnumbers
\startdata
C & -0.10$\pm$0.23 & \nodata  & \nodata  & -0.19$\pm$0.06 & \nodata  & \nodata  & 0.803 &\nodata  & \nodata  \\
O & -0.23$\pm$0.05 & \nodata & \nodata & -0.14$\pm$0.05 & \nodata  & \nodata  & 0.598 & \nodata  & \nodata  \\
Mg & -1.00$\pm$0.04 & -0.60$\pm$0.11 & -0.25$\pm$0.26 & -0.80$\pm$ 0.02 & -0.50$\pm$0.02 & -0.33$\pm$0.03 & 0.531 & 0.407 & 0.162 \\
Si & -1.14$\pm$0.06 & -1.11$\pm$0.12 & -1.05$\pm$0.09 & -0.57$\pm$0.03 & -0.68$\pm$0.03 & -0.36$\pm$0.02 & 0.305 & 0.247 & 0.129 \\
S & -0.879$\pm$ 0.28 & -1.02$\pm$ 0.10 & -0.87$\pm$ 0.14 & -0.091$\pm$ 0.04 & -0.31$\pm$ 0.02 & -0.02$\pm$ 0.04 & 0.290 & 0.137 & 0.106 \\
Ti & -2.05$\pm$0.06 & -1.48$\pm$0.15 & -1.45$\pm$0.09 & -1.96$\pm$0.03 & -1.63$\pm$0.02 & -1.23$\pm$0.02 & 0.430 & 0.401 & 0.189 \\
Cr & -1.45$\pm$0.06 & -1.18$\pm$0.08 & -1.33$\pm$0.16 & -1.51$\pm$0.06 & -1.13$\pm$0.02 & -0.93$\pm$0.02 & 0.470 & 0.368 & 0.155 \\
Fe & -1.28$\pm$0.04 & -1.28$\pm$0.04 & -1.28$\pm$0.07 & -1.51$\pm$0.03 & -1.51$\pm$0.03 & -1.18$\pm$0.02 & 0.437 & 0.437 & 0.181 \\
Ni & -1.49$\pm$0.06 & -1.29$\pm$0.08 & -1.41$\pm$0.14 & -1.83$\pm$0.04 & -1.26$\pm$0.02 & -1.11$\pm$0.02 & 0.599 & 0.338 & 0.141 \\
Cu & -0.71$\pm$0.09 & -1.15$\pm$0.42 & \nodata & -1.10$\pm$0.06 & -0.44$\pm$0.09 & \nodata & 0.711 & 0.325 & \nodata \\
Zn & -0.61$\pm$0.07 & -0.73$\pm$0.07 & -0.51$\pm$0.14 & -0.38$\pm$0.04 & -0.36$\pm$0.02 & -0.31$\pm$0.02 & 0.555 & 0.358 & 0.168 \\
\enddata
\end{deluxetable*}

\section{The collective behavior of depletions in the MW, LMC, and SMC} \label{deps_vs_fe}

\subsection{Constraints from MW, LMC, and SMC measurements}

\begin{figure*}
\centering
\includegraphics[width=\textwidth]{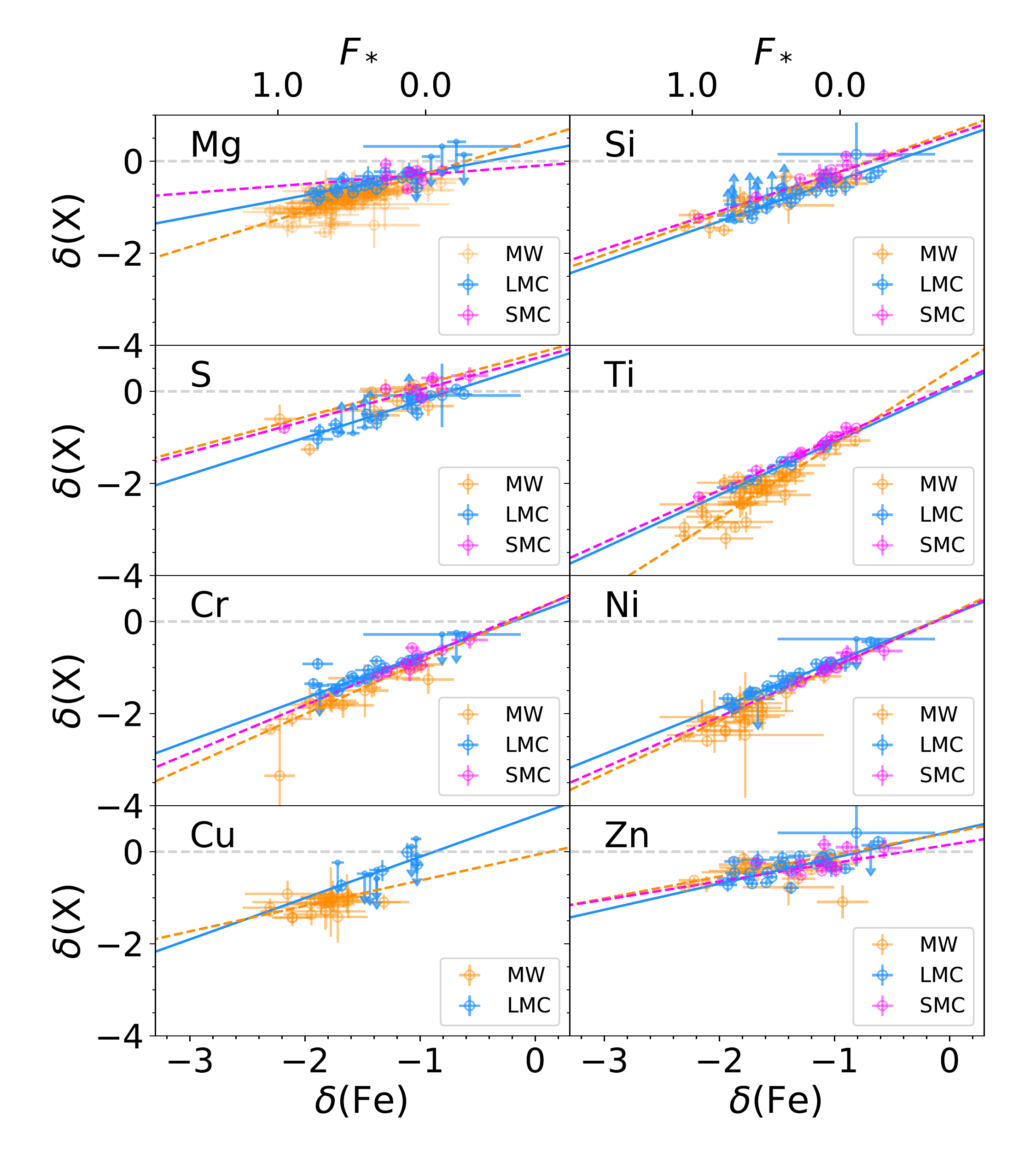}
\caption{Depletions (log fraction in the gas-phase) for Mg, Si, S, Cr, Ni, Cu, Zn, Ti as a function of iron depletions ($\delta$(Fe)). The fits of the depletions are shown by the colored lines (see Equation \ref{fstar_equation} with coefficients given in Table 2). Orange, blue, and magenta correspond to the Milky Way \citep{jenkins2009}, LMC \citep{RD2021}, and SMC \citep{jenkins2017}, respectively.}
\label{plot_deps_fe}
\end{figure*}

\indent In their large sample of depletions obtained in the Milky Way, \citet{jenkins2009} established that depletions for different elements tightly correlate with each other, and that the collective advancement of depletions for all elements can be described by the parameter $F_*$. In this section, we examine the relations between depletions of different elements in the Milky Way, LMC, and SMC. These relations are shown in Figure \ref{plot_deps_fe}, where both the depletion measurements for several elements are shown as a function of Fe depletions, in all three galaxies. Also shown are the fitted relations obtained from the $A_{\mathrm{X}}$, $B_{\mathrm{X}}$ and $z_{\mathrm{X}}$ coefficients (Table 2).\\
\indent We note that a vertical displacement of the relations $\delta$(X)---$\delta$(Fe) between the MW, LMC, and SMC could simply be due to differences and/or uncertainties on the assumed total abundances. However, differences in their slopes will reflect real differences in the rates at which elements deplete from the gas to the dust phase.\\
\indent For most elements (Si, S, Ni, Zn, Cr), the relation between depletions of X and Fe does not significantly change between the MW, LMC, and SMC. The invariance of the $\delta$(X) --- $\delta$(Fe) relation with metallicity, at least over the 0.2 --- 1 $Z_{\odot}$ range, implies that Fe, the interstellar abundance and depletion of which are straightforward to measure in UV spectra thanks to its numerous transitions with a range of oscillator strengths, can be used as a proxy for the depletions of other elements that are more difficult to measure.\\
\indent However, there are appreciable differences in the $\delta$(X)---$\delta$(Fe) relations between the MW, LMC, and SMC for Mg, Ti, and more marginally, Cu. Depletions of Cu are only measured for a small sample of sight-lines in the MW and LMC, with a high fraction of LMC sight-lines only having upper limits. As a result, the relatively large difference in the $\delta$(Cu)---$\delta$(Fe) between the MW and LMC is not statistically significant (1$\sigma$ difference). However, for Mg, the slopes differ at the 4$\sigma$ level between the MW and the LMC, and at the 3$\sigma$ level between the MW and the SMC, a decrease in the steepness of the relation with decreasing metallicity that is clearly seen in Figure \ref{plot_deps_fe}. For Ti, the LMC and SMC relations are in almost prefect agreement, but differ from the MW relation at the 3$\sigma$ level. Mn was not measured in the LMC as part of METAL, but \citet{jenkins2017} report significant differences between the rate of depletion of Mn in the SMC and MW. To explain differences in the depletion rates between different galaxies, \citet{jenkins2017} offered the conjecture that the lower abundance of C in the SMC and the consequences of the mix of PAH and silicates could influence the chemical affinities of various atoms to dust and hence their respective rates of depletions. \\

\subsection{Abundance ratios in the MW, LMC, and SMC}\label{abundance_ratio_section}

 \begin{figure*}
 \centering
\includegraphics[width=8cm]{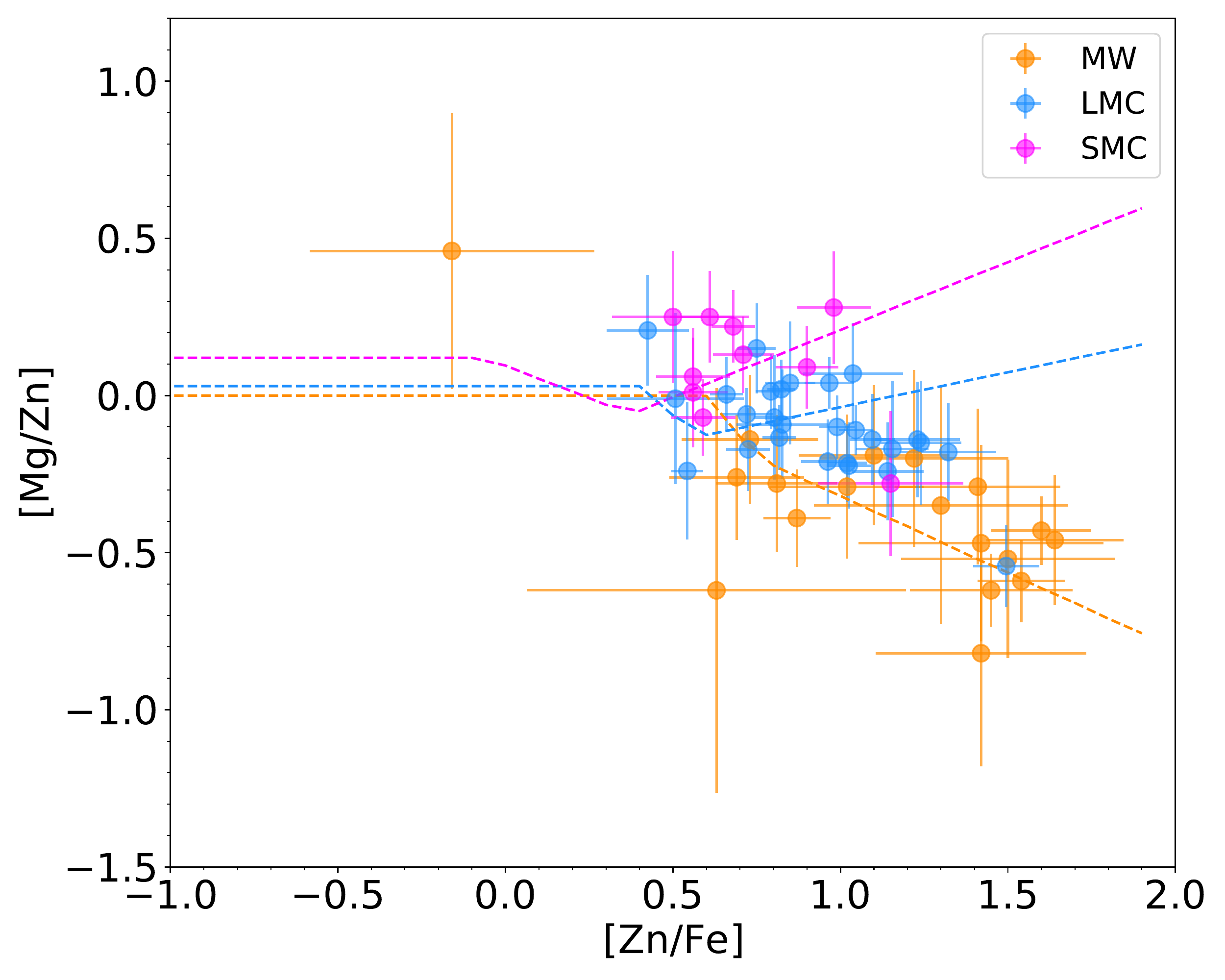}
\includegraphics[width=8cm]{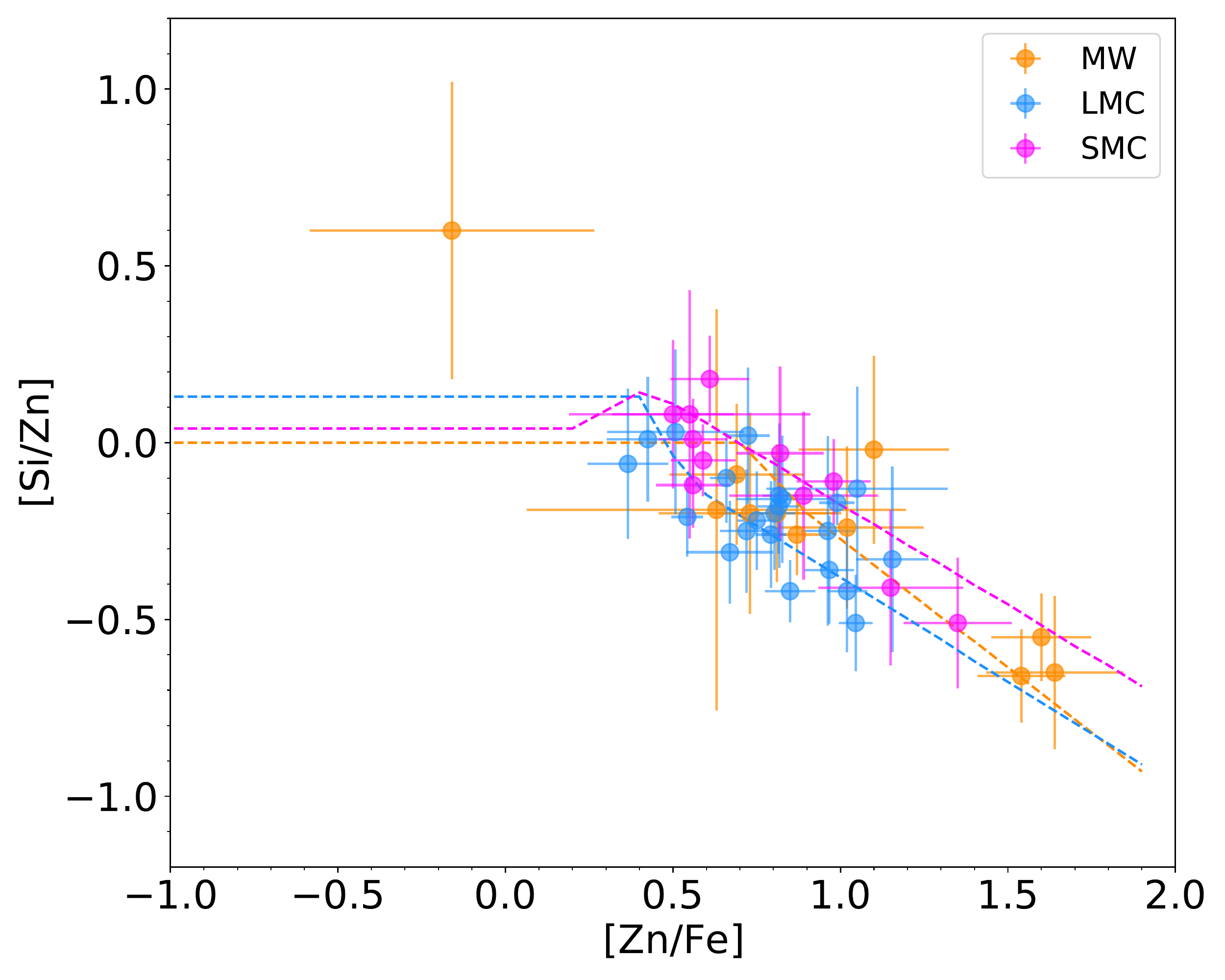}
\includegraphics[width=8cm]{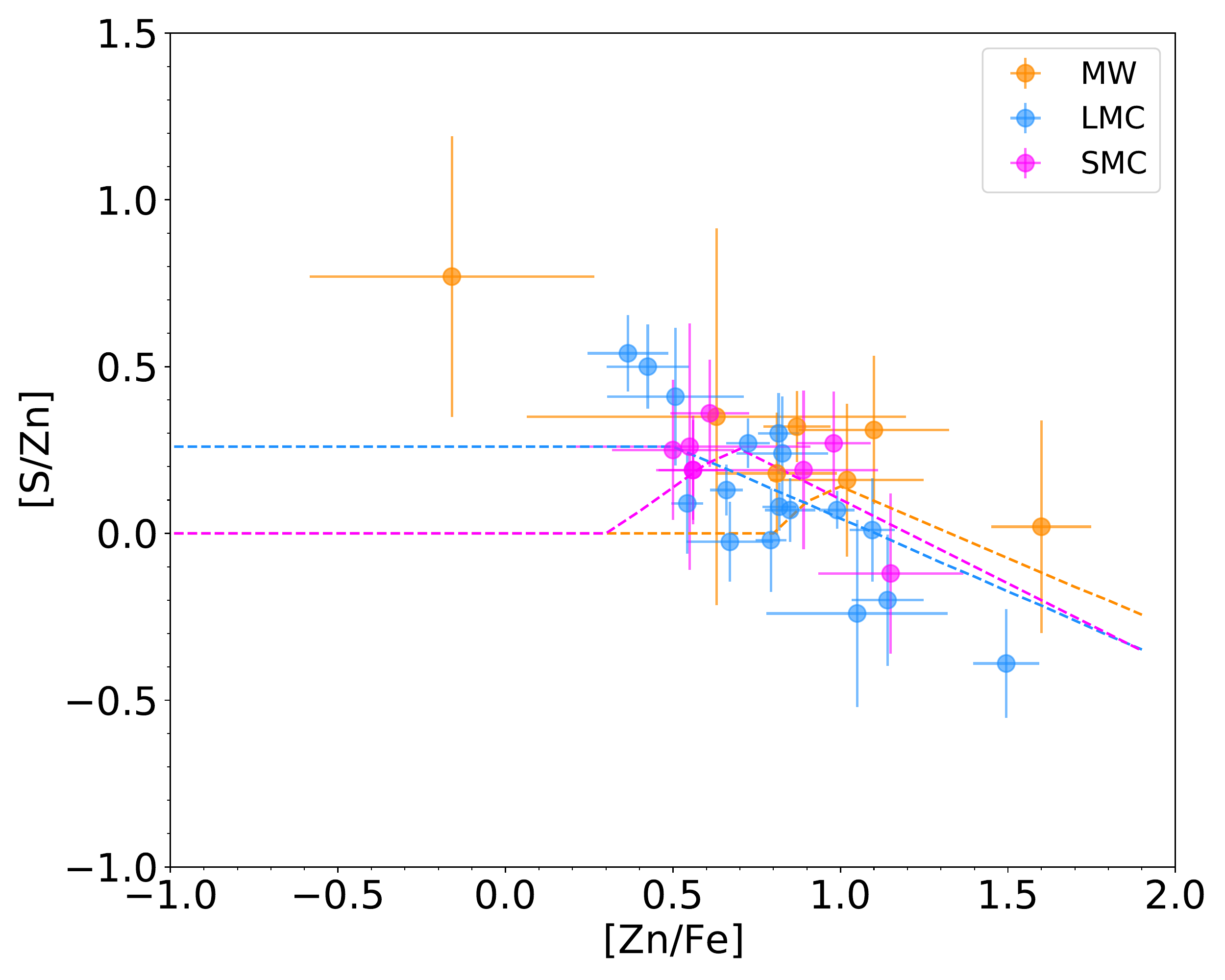}
\includegraphics[width=8cm]{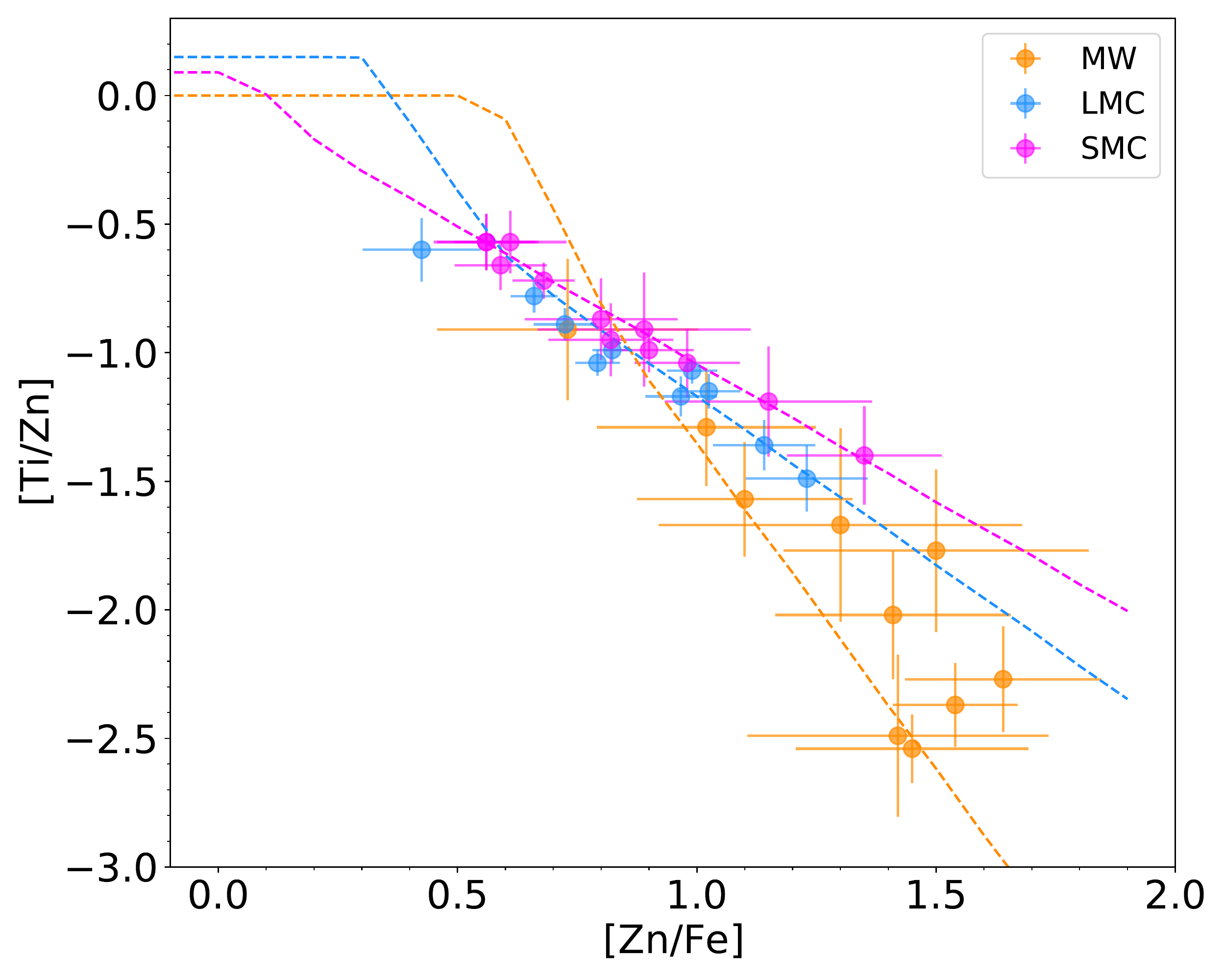}
\includegraphics[width=8cm]{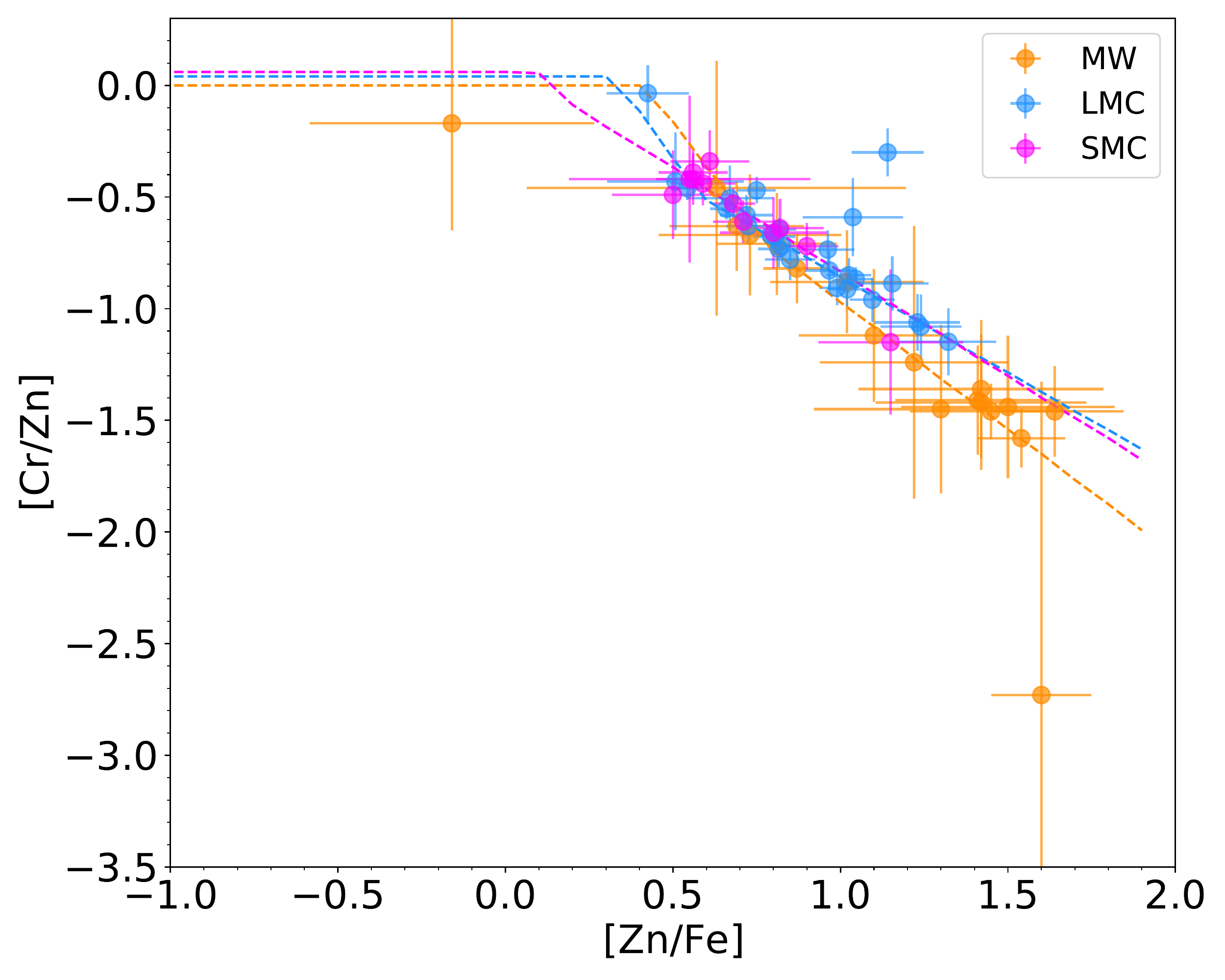}
\includegraphics[width=8cm]{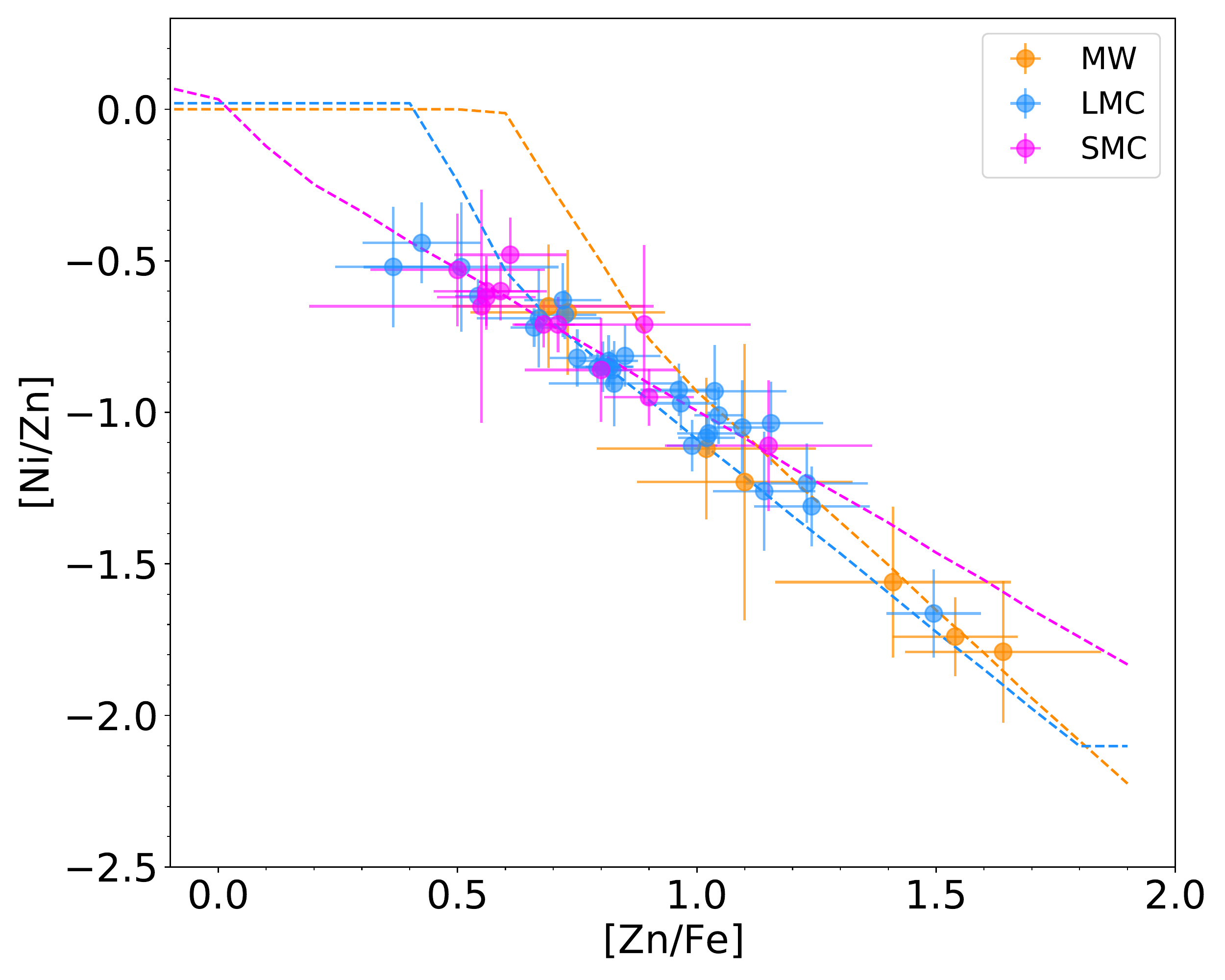}
\caption{Relation between abundance ratios [X/Zn] with X = Mg, Si, S, Cr in the MW (orange), LMC (blue), and SMC (magenta). The dashed lines are obtained from the relation between depletions of different elements in the MW, LMC, and SMC (see Equation \ref{fstar_equation}, Table 2, and Figure \ref{plot_deps_fe})}
\label{plot_abundance_ratios_Zn}
\end{figure*}

 \begin{figure*}
 \centering
\includegraphics[width=8cm]{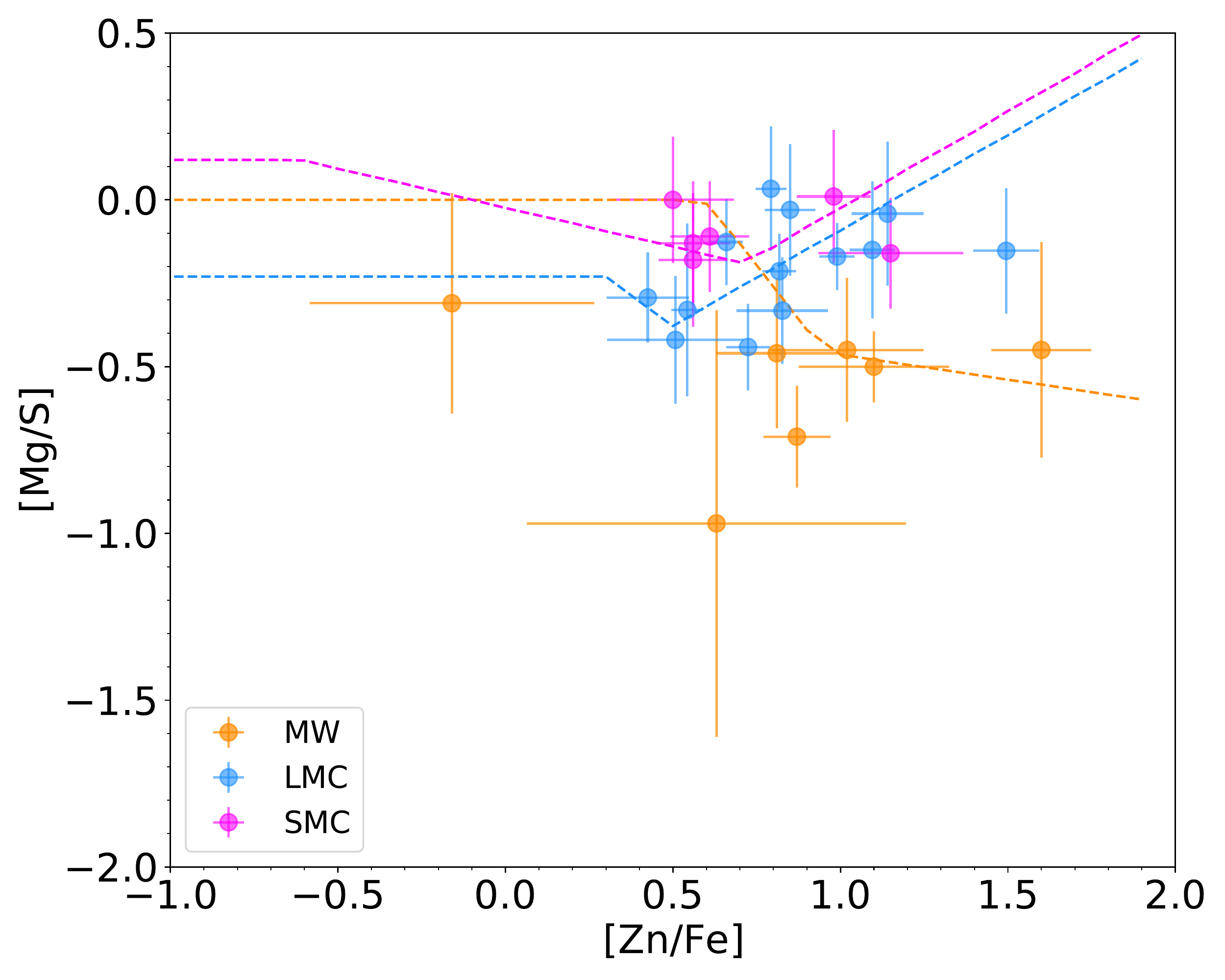}
\includegraphics[width=8cm]{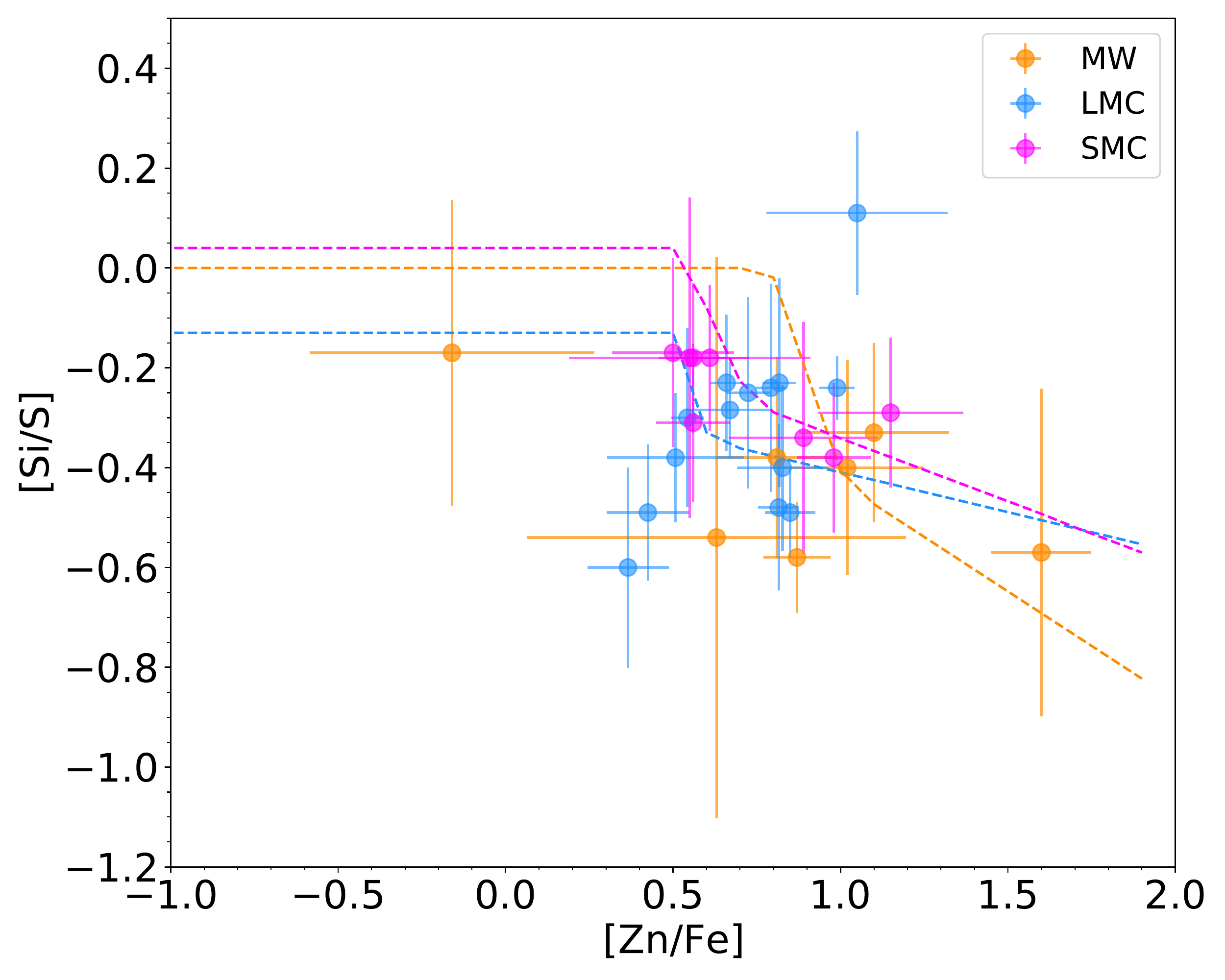}
\includegraphics[width=8cm]{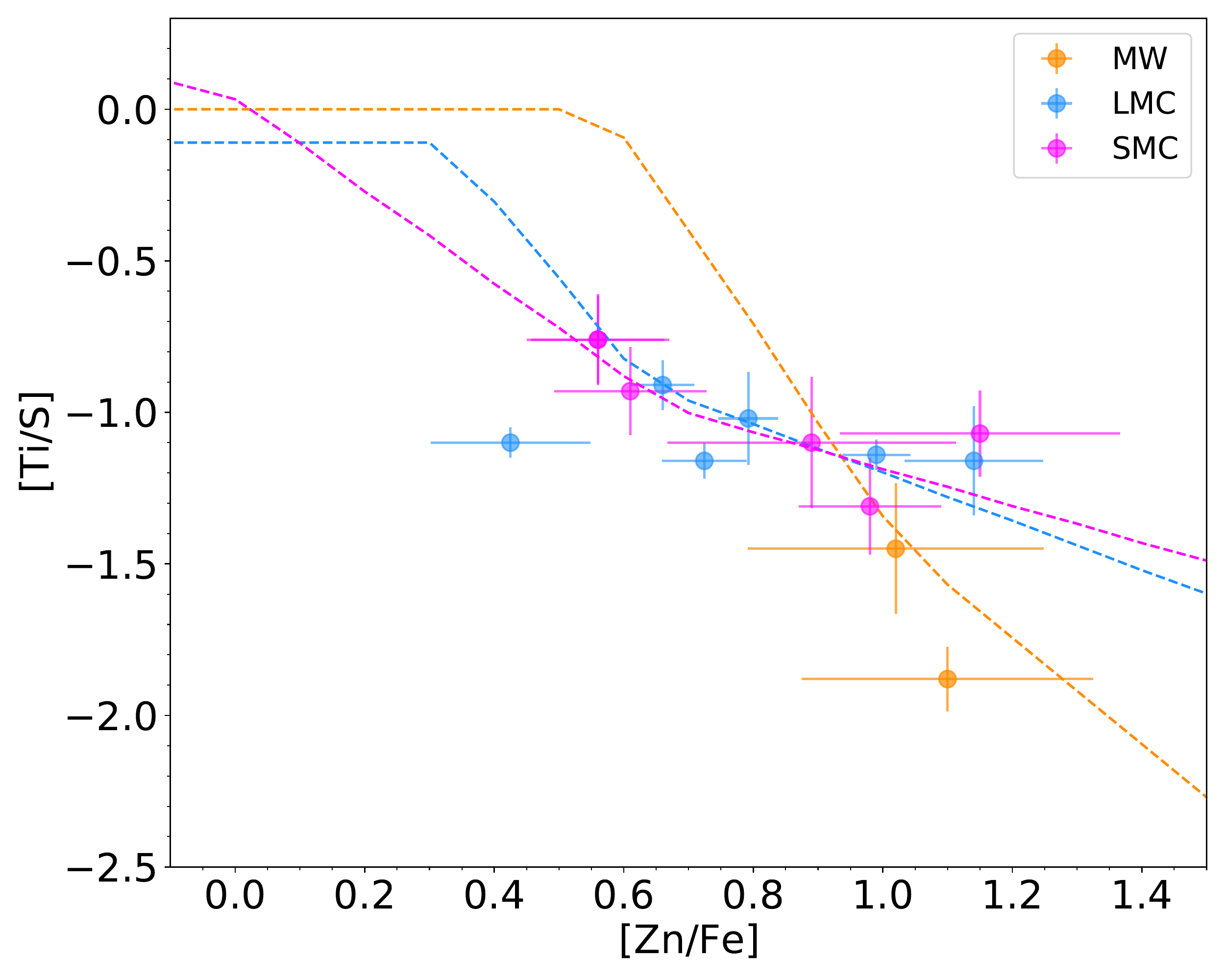}
\includegraphics[width=8cm]{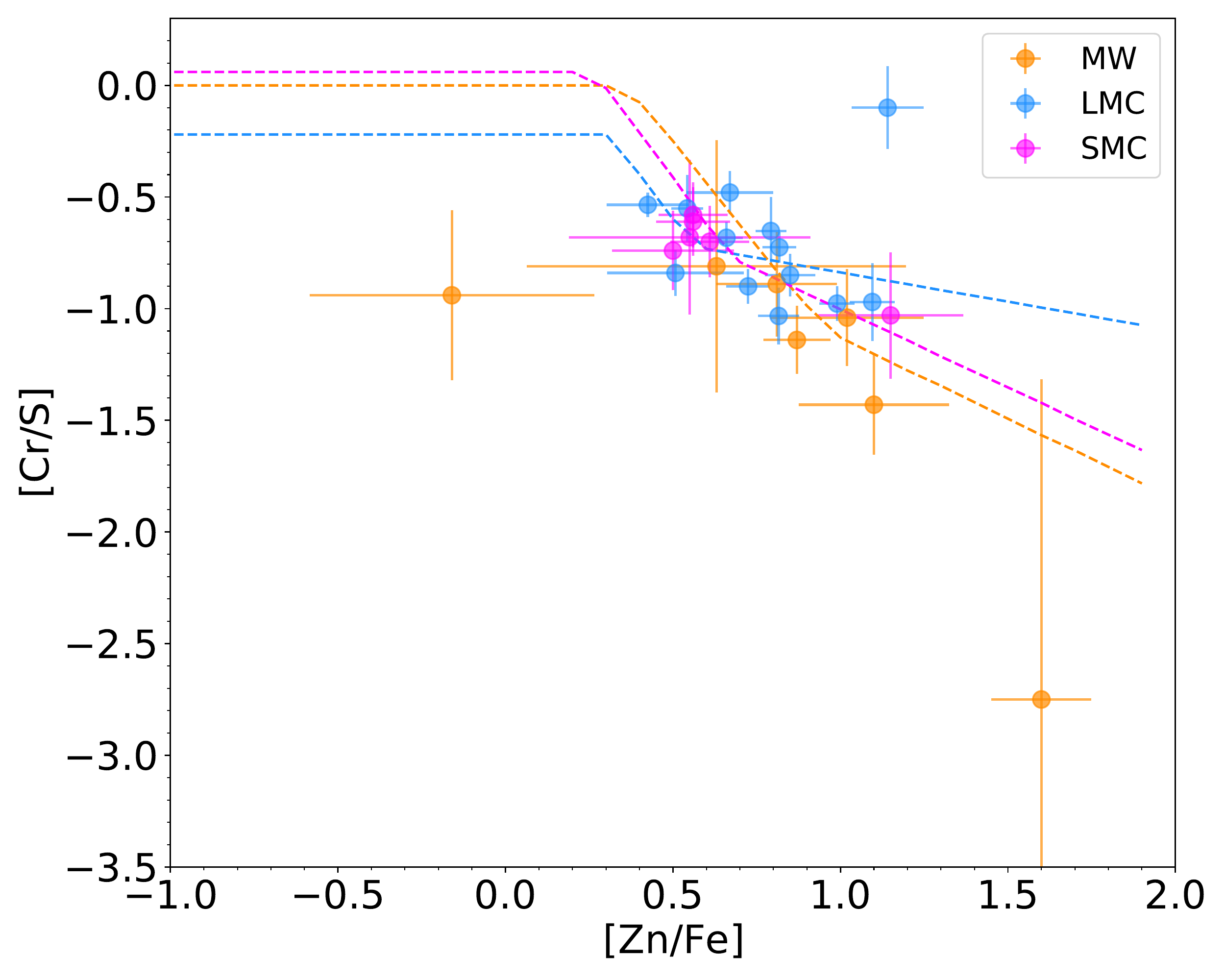}
\includegraphics[width=8cm]{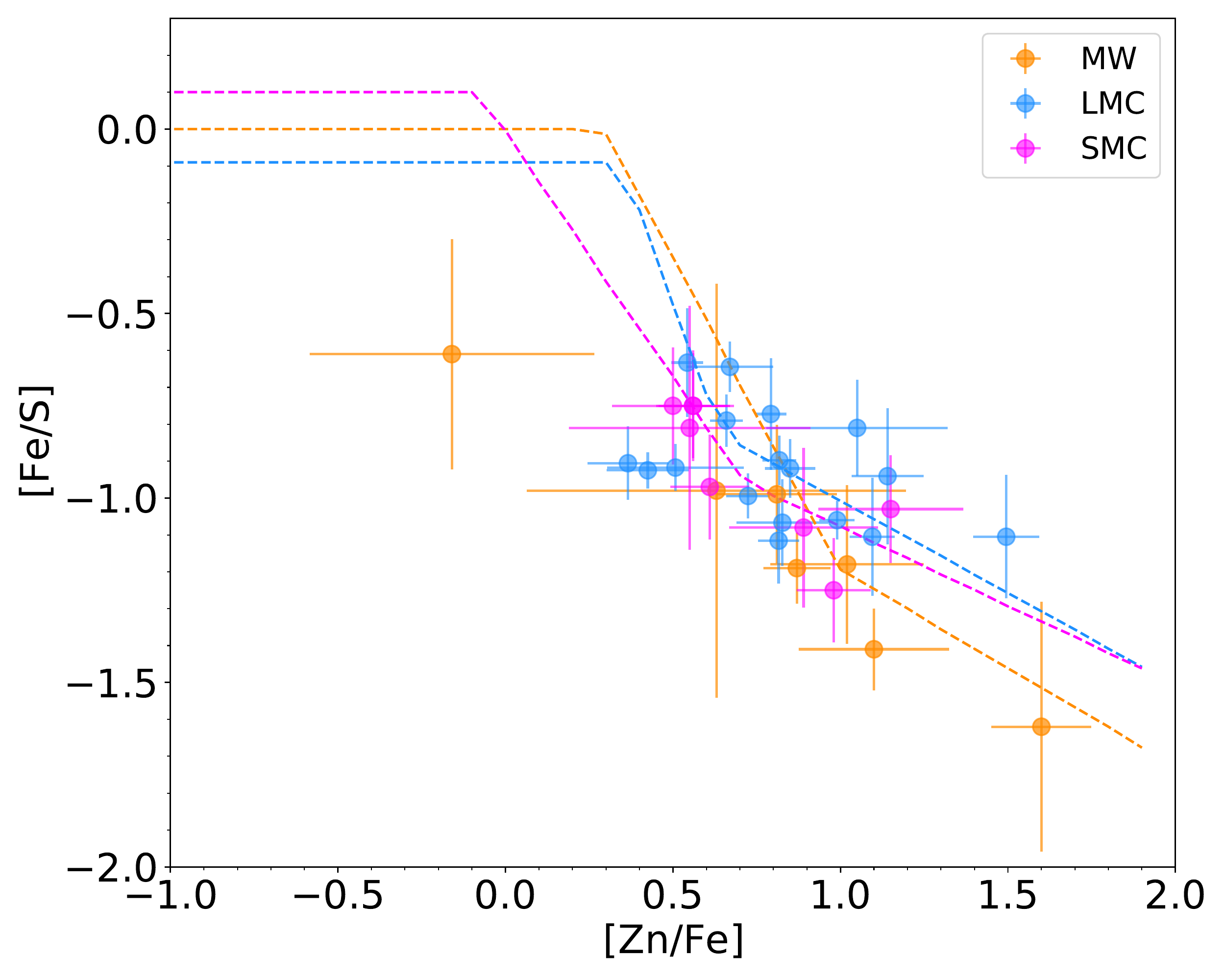}
\includegraphics[width=8cm]{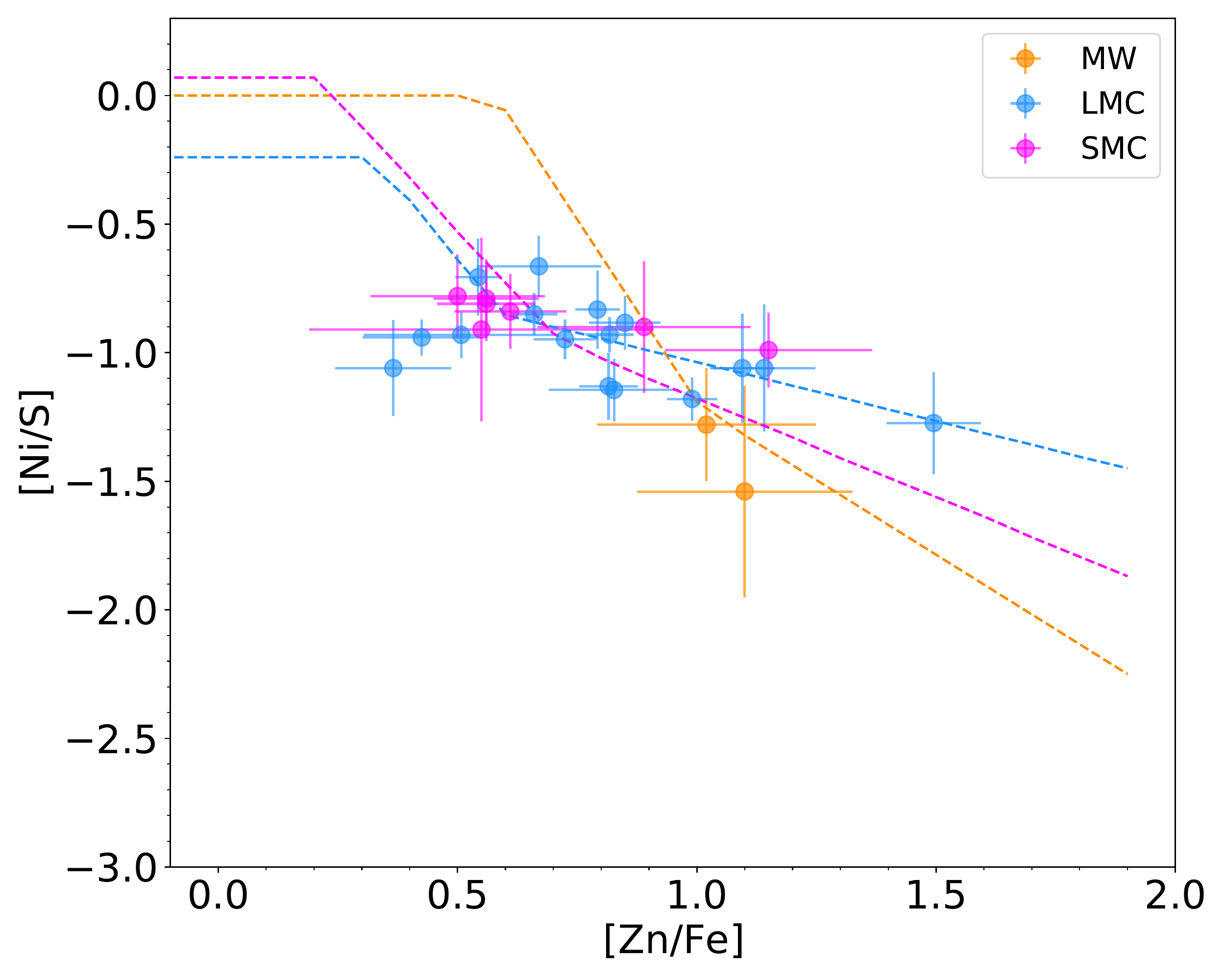}
\caption{Relation between abundance ratios [X/S] with X = Mg, Si, Fe, Cr in the MW (orange), LMC (blue), and SMC (magenta). The dashed lines are obtained from the relation between depletions of different elements in the MW, LMC, and SMC (see Equation \ref{fstar_equation}, Table 2, and Figure \ref{plot_deps_fe})}
\label{plot_abundance_ratios_S}
\end{figure*}

\indent In nearby galaxies, total (gas+dust) abundances in the ISM can be estimated indirectly from the photospheric abundances of young stars recently formed out of the ISM. In more distant systems such as DLAs, depletions can only be inferred from abundance ratios of volatile to refractory elements, which are heavily influenced by their different rates of depletion. Thus, comparing the relations between different abundance ratios in the MW, LMC, and SMC can reveal key information about the depletion process in both local galaxies and more distant systems.\\
\indent We examine the relation between abundance ratios involving volatile elements Zn and S in the MW, LMC, and SMC. This is similar to the comparison performed by \citet{decia2018b}, but with the addition of the new LMC sample obtained from the METAL program. In Figures \ref{plot_abundance_ratios_S} and \ref{plot_abundance_ratios_Zn}, we show various abundance ratios ([X/Zn] and [X/S], respectively, for X = Mg, Si, S, Ti, Cr, Fe, Ni) in the MW, LMC, and SMC as a function of [Zn/Fe], which is commonly used as a depletion tracer in DLAs. For the MW, LMC, and SMC, both the abundance ratios toward individual sight-lines and the relations derived from the $A$, $B$, and $z$ coefficients relating depletions and $F_*$ (and hence relating depletions of different elements) are shown in Figures \ref{plot_abundance_ratios_Zn} and \ref{plot_abundance_ratios_S}. We note that, while plotted as independent (orthogonal) in x and y, error bars in Figures \ref{plot_abundance_ratios_Zn} and \ref{plot_abundance_ratios_S} should not be orthogonal when the same element is involved in both axes. \\
\indent Because depletions cannot exceed the zero value, the relations between abundance ratios derived from the $A$, $B$, and $z$ coefficients are not simple linear functions. The abundance ratio [X/Y] is determined by the depletions and $\alpha$-enhancement of X and Y via

\begin{equation}
\left [ \frac{\mathrm{X}}{\mathrm{Y}} \right ] = \delta(\mathrm{X}) - \delta(\mathrm{Y})  + \alpha(\mathrm{X}) - \alpha(\mathrm{Y})
\end{equation}

\noindent where $\alpha$(X) is the over- or under-abundance of X with respect to Fe relative to the solar (X/Fe)$_{\odot}$ ratio ($\alpha$(X) $=$ [X/Fe], where [X/Fe] is measured in stars). We note that $\alpha$(X) refers to abundance variations relative to solar in stars, where dust depletion effects are not occurring. In particular, $\alpha$(X) accounts for nucleosynthetic effects in $\alpha$-elements (e.g., Si, S, Mg), the so-called $\alpha$-enhancement. $\alpha$(X) is known from stellar abundances in the LMC and SMC listed in Table 1 ($\alpha$(X) $=$ 0 in the MW by construction). The depletion of X is $\delta$(X) $=$ $A_{\mathrm{X}}$($F_*$ $-$ $z_{\mathrm{X}}$) $+$ $B_{\mathrm{X}}$. $\delta$(X) is capped at zero value. This results in [X/Y] following the piece-wise linear functions shown in Figures \ref{plot_abundance_ratios_Zn} and \ref{plot_abundance_ratios_S}  in the MW, LMC, and SMC.\\
\indent The abundance ratios shown in Figures \ref{plot_abundance_ratios_Zn} and \ref{plot_abundance_ratios_S} are generally in reasonable agreement between the MW, LMC, and SMC. This is not surprising in light of Figure \ref{plot_deps_fe}, since abundance ratios depend primarily on the relative depletions of the elements involved, and Figure \ref{plot_deps_fe} shows that the relation between the depletions of different elements remains relatively constant between the MW, LMC, and SMC. The invariance of the relation between abundance ratios between the MW, LMC, and SMC implies that abundance ratios should in principle serve as accurate tracers of depletions in DLAs, at least down to the 20\% solar metallicity probed by local studies. Deriving calibrations of abundance ratios, in particular [Zn/Fe], in the MW, LMC and SMC to be used by the DLA community will be the subject of the METAL IV paper (Roman-Duval et al., in prep). \\
\indent As in Figure \ref{plot_deps_fe}, Mg (in the SMC) and Ti (in the MW) are the only exceptions to the invariance of abundance ratio variations between the MW, LMC, and SMC, deviating slightly from the other trends. For Mg, some of the rather large differences seen between the MW and the SMC (and to some extent the LMC) may be due to the limited dynamic range and small sample in depletion measurements outside our Galaxy, resulting in the diverging extrapolation of the fit at high [Zn/Fe] for the LMC and SMC.

\subsection{Estimating C and O depletions in the LMC and SMC}\label{estimating_c_and_o}

\indent The full suite of elements that make up most of the dust mass does not necessarily have depletion measurements in all three galaxies. In particular, C and O in the LMC and SMC are not measured because the UV transitions of C and O are either too saturated or too weak. Unfortunately, C and O constitute the largest mass reservoir of heavy elements that can be included in dust. This limitation can be circumvented thanks to the relative invariance of the collective behavior of depletions observed in the MW, LMC, and SMC. Here as in \citet[][their Section 7]{RD2021}, we therefore use the assumption hat the relation between C or O depletions and iron depletions behaves similarly in the Milky Way, LMC, and SMC, as is the case for other elements. Knowing the iron depletions for all our sight lines in all 3 galaxies, we apply the known MW relation between $\delta$(C) or $\delta$(O) and $\delta$(Fe) (Equation \ref{fstar_equation} and coefficients in Table 2) from \citep{jenkins2009} to obtain an estimate of $\delta$(C) or $\delta$(O) for each sight line. The error on the $A_{\mathrm{X}}$ and $B_{\mathrm{X}}$ coefficients are propagated through the calculation of C and O depletions. \\
\indent We note that a deficiency of carbon relative to other elements in the LMC (log C/O = $-$0.56 in the LMC versus $-$0.30 in the MW, and $-$0.62 in the SMC) may potentially affect the rate of carbon depletions compared to those of other elements. For example, the fraction of carbonaceous dust and PAHs relative to silicates is different between the MW, LMC, and SMC \citep{chastenet2019}, which could be attributed to the different chemical affinities of dust compounds induced by the lower carbon abundance in the LMC and SMC compared to the MW. \\

\section{The dust-to-gas and dust-to-metal ratios}\label{computing_dh}

\indent With depletions in hand for a suite of elements in the MW, LMC, and SMC, we can derive the dust-to-gas and dust-to-metal ratios toward each sight-line by summing the mass weighted dust-phase abundance of all elements:

\begin{equation}\label{doh_equation}
D/G = \frac{1}{1.36} \sum_{X} \left (1-10^{\delta(X)} \right ) \left (\frac{N(X)}{N_\mathrm{H}} \right )_{\mathrm{tot}} W(X)
\label{dg_eq}
\end{equation}

\noindent and, 

\begin{equation}
D/M = \frac{\sum_{X} \left (1-10^{\delta(X)} \right ) \left (\frac{N(X)}{N_{\mathrm{H}}} \right )_{\mathrm{tot}} W(X)}{\sum_{X}  \left (\frac{N(X)}{N_{\mathrm{H}}} \right )_{\mathrm{tot}} W(X)}
\label{dm_eq}
\end{equation}

\noindent where (N(X)/N$_{\mathrm{H}})_{\mathrm{tot}}$ is the total abundance of element X in the galaxy (MW, LMC, or SMC), assumed to be that of stellar photospheres of young stars, and $W(X)$ is the atomic weight of element X. Throughout this paper, when depletions are estimated from the $A_{\mathrm{X}}$, $B_{\mathrm{X}}$, and $z_{\mathrm{X}}$ coefficients, errors on these coefficients are propagated, for example through the calculation of D/G and D/M.

\section{The relation between depletions, D/M, D/G and hydrogen column density in the MW, LMC, and SMC} \label{section_nh}

\subsection{Relation between depletions and hydrogen column density}\label{deps_vs_lognh}

\begin{figure*}
\centering
\includegraphics[width=\textwidth]{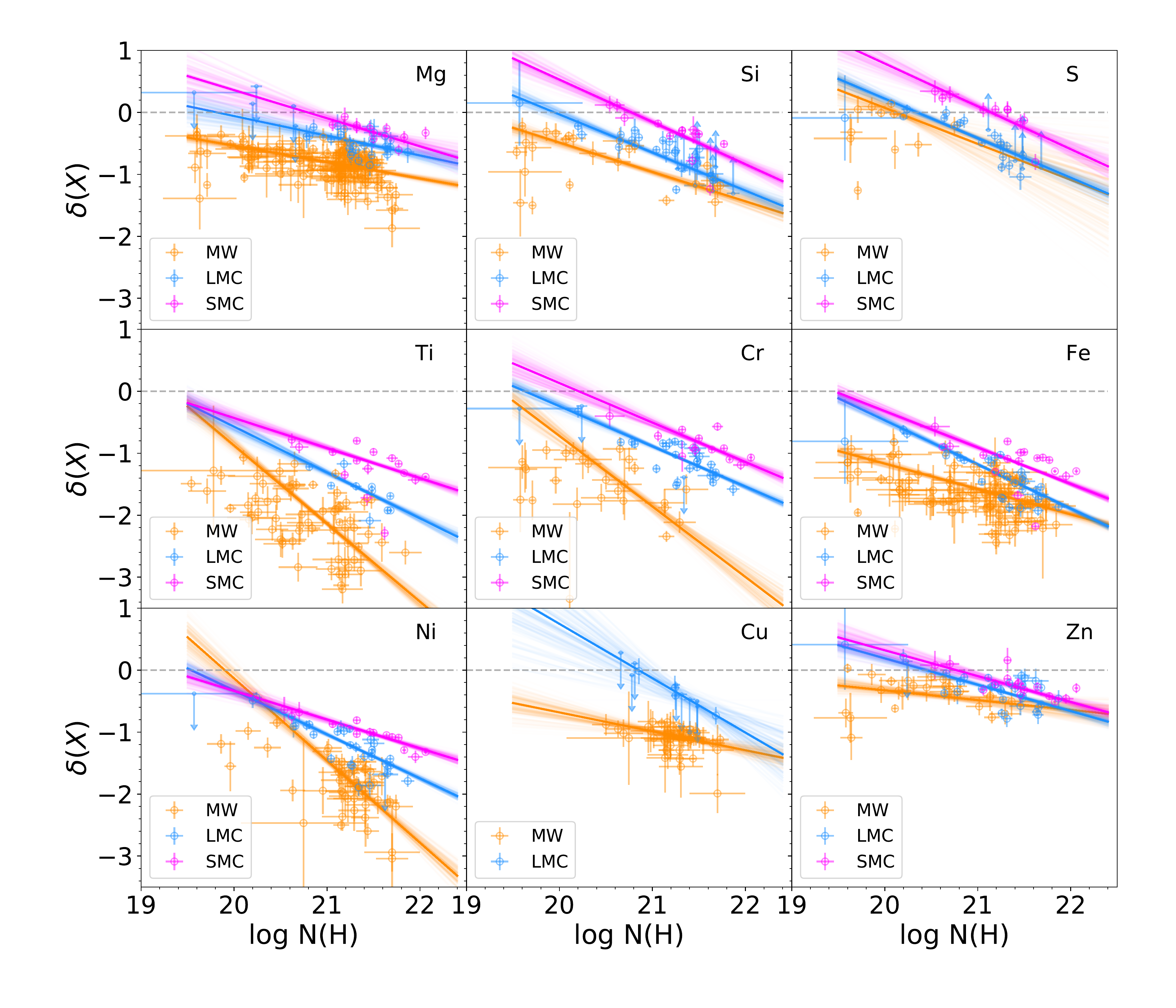}
\caption{Depletions (log fraction in the gas-phase) for Mg, Si, S, Cr, Fe, Ni, Cu, Zn, Ti, as a function of $\log$ N(H). Orange, blue, and magenta correspond to the Milky Way \citep{jenkins2009}, LMC \citep{RD2021}, and SMC \citep{jenkins2017}, respectively. The fraction of metals in the gas-phase decreases with increasing column density. Different realizations of the fits are shown with transparencies computed as the square root of the likelihood of each realization. }
\label{plot_deps_nh}
\end{figure*}

\indent Based on the METAL spectra, \citet{RD2021} investigated the physical parameters driving the depletion levels in the LMC, and found that the hydrogen column density was the primary parameter correlated with the depletions of different elements. In a face-on disk galaxy such as the LMC, this is consistent with the gas density being a determining factor in the depletion levels. Indeed, variations in the column density stem from either variation in the density, or in the path length. The latter result from variations of the scale height of the gas perpendicular to the plane of the LMC, the magnitude of which are smaller compared to variations in the density. Therefore, the hydrogen column density should be a direct tracer of the average gas density along the line-of-sight. Of course, the average density itself will result from gas at a range of densities where depletions occur at correspondingly varying rates, and our measurements therefore reflect the average depletion behavior in the interstellar gas density structure.\\
\indent In this section, we investigate whether depletion measurements in the MW and SMC also support the gas density and column density being the main parameters driving depletion levels in the ISM, and if so, whether the trends between depletions and $\log$ N(H) are consistent with those observed in the LMC.\\
\indent The depletions for all elements measured in the MW, LMC, and SMC are shown as a function of $\log$ N(H) in Figure \ref{plot_deps_nh}. For all galaxies and elements, there is a clear anti-correlation between $\delta$(X) and $\log$ N(H): as the column density increases, an increasing fraction of metals are locked into dust grains (recall that depletions correspond to the log fractions of metals in the gas). As a result, even for volatile elements such as S and Zn often used as metallicity tracers in DLAs, the depletion level can exceed $-$0.5 dex for $\log$ N(H) $>$ 21 cm$^{-2}$, even at 20\% solar metallicity.\\
\indent We fit the relation between depletions and $\log$ N(H) using linear functions, as in \citet{RD2021}:

\begin{equation}\label{dep_log_nh_eq}
\delta(\mathrm{X}) = B_{\mathrm{H}}(\mathrm{X}) + A_{\mathrm{H}}(\mathrm{X}) \left (\log N(\mathrm{H}) - \log N_{\mathrm{H}_0}(\mathrm{X}) \right )
\end{equation}

\noindent which has the same form as Equation \ref{fstar_equation}, but is applicable to N(H) instead of $F_*$. The parameter $N_{\mathrm{H}_0}(\mathrm{X}$) is introduced to remove the covariance between errors on the slope, $A_{\mathrm{H}}$(X), and intercept, $B_{\mathrm{H}}$(X), of the relation, and is given by:

\begin{equation}
\log N_{\mathrm{H}_0}(\mathrm{X}) = \frac{\sum_{\mathrm{los}}\frac{\log N(H)}{\sigma(\delta(X))^2}}{\sum_{\mathrm{los}}\frac{1}{\sigma(\delta(X))^2}}
\end{equation}

\indent The resulting $A_{\mathrm{H}}$(X),  $B_{\mathrm{H}}$(X), and N$_{\mathrm{H}_0}$ for the MW, LMC, and SMC are listed in Table 3 and shown in Figure \ref{plot_deps_nh}. The coefficients for the LMC were computed in \citet{RD2021} and are repeated here for easy comparison with the MW and SMC. For the MW, where depletions are measured for sight-lines with $\log$ N(H) down to 18 cm$^{-2}$, we only fit $\delta$(X) vs. $\log$ N(H) in the range $\log$ N(H) $=$ 20---22 cm$^{-2}$ so as to be consistent with the LMC and SMC, where only sight-lines with $\log$ N(H) $>$ 20 cm$^{-2}$ were measured. Pre-computed values of depletions for $\log$ N(H) $=$ 20, 21, and 22 cm$^{-2}$ in the MW, LMC, and SMC are listed in Table 4 for convenience. Those values are computed from Equation \ref{dep_log_nh_eq} and the coefficients listed in Table 3. \\
\indent In Table 4, we also list estimates of C and O depletions obtained using the approach described in Section \ref{estimating_c_and_o}, which relies on the invariance of the collective behavior of depletions between the MW, LMC, and SMC. For a given $\log$ N(H), we use Equation \ref{dep_log_nh_eq} and the coefficients listed in Table 3 to estimate the Fe depletion. We then make use of the MW relation between the depletions of Fe and C or O (Equation \ref{fstar_equation} and coefficients in Table 2) to estimate $\delta$(C) and $\delta$(O) in the LMC and SMC. This allows us to compute the D/M in all three galaxies, from the depletions listed in Table 4 and Equation \ref{dm_eq}.

\begin{deluxetable*}{cccc|ccc|ccc}
\tablenum{3}
\tablecaption{$A_{\mathrm{H}}$(X), $B_{\mathrm{H}}$(X), and $N_{\mathrm{H}_0}$(X) coefficients relating depletions and $\log$ N(H) in the MW, LMC, and SMC \label{tab:dep-lognh}}
\tablewidth{0pt}
\tablehead{
\colhead{Element} &  \multicolumn{3}{c}{$A_{\mathrm{H}}$(X)} & \multicolumn{3}{c}{$B_{\mathrm{H}}$(X)} & \multicolumn{3}{c}{$\log$ $N_{\mathrm{H}_0}$(X) }\\
\cline{2-4}  \cline{5-7} \cline{8-10}
& \colhead{MW} & \colhead{LMC} & \colhead{SMC} & \colhead{MW} & \colhead{LMC} & \colhead{SMC} & \colhead{MW} & \colhead{LMC} &\colhead{SMC} 
}
%\decimalcolnumbers
\startdata
C & 0.12$\pm$ 0.12 & \nodata & \nodata  & -0.14$\pm$ 0.04 & \nodata & \nodata & 21.340 & \nodata  & \nodata  \\
O & -0.03$\pm$ 0.05 & \nodata  & \nodata & -0.15$\pm$ 0.01 & \nodata   & \nodata & 21.265 & \nodata & \nodata  \\
Mg & -0.26$\pm$ 0.03 & -0.32$\pm$ 0.09 & -0.45$\pm$ 0.14 & -0.83$\pm$ 0.01 & -0.50$\pm$ 0.02 & -0.32$\pm$ 0.04 & 21.121 & 21.374 & 21.490 \\
Si & -0.47$\pm$ 0.05 & -0.61$\pm$ 0.07 & -0.68$\pm$ 0.08 & -0.81$\pm$ 0.03 & -0.67$\pm$ 0.03 & -0.36$\pm$ 0.03 & 20.682 & 21.040 & 21.308 \\
S & -0.58$\pm$ 0.22 & -0.64$\pm$ 0.06 & -0.69$\pm$ 0.12 & -0.03$\pm$ 0.05 & -0.32$\pm$ 0.02 & -0.02$\pm$ 0.04 & 20.171 & 20.842 & 21.166 \\
Ti & -1.26$\pm$ 0.08 & -0.74$\pm$ 0.07 & -0.48$\pm$ 0.06 & -2.03$\pm$ 0.03 & -1.63$\pm$ 0.02 & -1.15$\pm$ 0.02 & 20.916 & 21.427 & 21.490 \\
Cr & -1.14$\pm$ 0.13 & -0.65$\pm$ 0.04 & -0.64$\pm$ 0.09 & -1.60$\pm$ 0.05 & -1.14$\pm$ 0.01 & -0.90$\pm$ 0.02 & 20.769 & 21.383 & 21.618 \\
Fe & -0.41$\pm$ 0.03 & -0.71$\pm$ 0.03 & -0.59$\pm$ 0.04 & -1.62$\pm$ 0.02 & -1.39$\pm$ 0.01 & -1.18$\pm$ 0.01 & 21.097 & 21.288 & 21.469 \\
Ni & -1.33$\pm$ 0.11 & -0.71$\pm$ 0.04 & -0.46$\pm$ 0.06 & -1.77$\pm$ 0.03 & -1.26$\pm$ 0.01 & -1.04$\pm$ 0.02 & 21.232 & 21.318 & 21.527 \\
Cu & -0.30$\pm$ 0.10 & -0.88$\pm$ 0.30 &\nodata  & -1.08$\pm$ 0.02 & -0.44$\pm$ 0.09 & \nodata  & 21.293 & 21.354 & \nodata  \\
Zn & -0.16$\pm$ 0.07 & -0.43$\pm$ 0.04 & -0.42$\pm$ 0.07 & -0.42$\pm$ 0.03 & -0.36$\pm$ 0.01 & -0.32$\pm$ 0.02 & 20.588 & 21.299 & 21.543 \\
\enddata
\end{deluxetable*}

\begin{deluxetable*}{cccc|ccc|ccc}
\tablenum{4}
\tablecaption{Depletion and D/M values obtained from linear fits to hydrogen column density for $\log$ N(H) $=$ 20, 21, and 22 cm$^{-2}$ \label{tab:dep-lognh-summary}}
\tablewidth{0pt}
\tablehead{
\colhead{} &  \multicolumn{3}{c}{$\log$ N(H) $=$ 20 cm$^{-2}$} & \multicolumn{3}{c}{$\log$ N(H) $=$ 21 cm$^{-2}$} & \multicolumn{3}{c}{$\log$ N(H) $=$ 22 cm$^{-2}$} \\
\cline{2-4}  \cline{5-7} \cline{8-10}
& \colhead{MW} & \colhead{LMC} & \colhead{SMC} & \colhead{MW} & \colhead{LMC} & \colhead{SMC} & \colhead{MW} & \colhead{LMC} &\colhead{SMC} 
}
\startdata
$\delta$(C) & -0.13$\pm$0.16 & -0.07$\pm$0.28 & -0.06$\pm$0.30 & -0.16$\pm$0.09 & -0.13$\pm$0.15 & -0.11$\pm$0.19 & -0.19$\pm$0.06 & -0.19$\pm$0.06 & -0.16$\pm$0.10 \\
$\delta$(O) & -0.05$\pm$0.06 & 0.00$\pm$0.07 & 0.00$\pm$0.08 & -0.12$\pm$0.05 & -0.05$\pm$0.06 & -0.00$\pm$0.06 & -0.19$\pm$0.05 & -0.18$\pm$0.05 & -0.11$\pm$0.05 \\
$\delta$(Mg) & -0.54$\pm$0.04 & -0.06$\pm$0.12 & 0.00$\pm$0.21 & -0.80$\pm$0.01 & -0.38$\pm$0.04 & -0.09$\pm$0.08 & -1.07$\pm$0.03 & -0.70$\pm$0.06 & -0.55$\pm$0.08 \\
$\delta$(Si) & -0.49$\pm$0.05 & -0.03$\pm$0.08 & 0.00$\pm$0.11 & -0.96$\pm$0.04 & -0.65$\pm$0.03 & -0.15$\pm$0.04 & -1.44$\pm$0.08 & -1.26$\pm$0.07 & -0.84$\pm$0.06 \\
$\delta$(S) & 0.00$\pm$0.06 & 0.00$\pm$0.05 & 0.00$\pm$0.15 & -0.51$\pm$0.19 & -0.42$\pm$0.02 & 0.00$\pm$0.05 & -1.09$\pm$0.41 & -1.05$\pm$0.07 & -0.59$\pm$0.11 \\
$\delta$(Ti) & -0.88$\pm$0.07 & -0.57$\pm$0.09 & -0.43$\pm$0.08 & -2.14$\pm$0.03 & -1.31$\pm$0.03 & -0.92$\pm$0.03 & -3.40$\pm$0.09 & -2.05$\pm$0.04 & -1.40$\pm$0.03 \\
$\delta$(Cr) & -0.72$\pm$0.11 & -0.24$\pm$0.06 & 0.00$\pm$0.14 & -1.86$\pm$0.06 & -0.89$\pm$0.02 & -0.50$\pm$0.06 & -3.00$\pm$0.16 & -1.54$\pm$0.03 & -1.14$\pm$0.04 \\
$\delta$(Fe) & -1.17$\pm$0.04 & -0.47$\pm$0.04 & -0.32$\pm$0.06 & -1.58$\pm$0.02 & -1.18$\pm$0.01 & -0.91$\pm$0.02 & -1.99$\pm$0.03 & -1.89$\pm$0.02 & -1.49$\pm$0.03 \\
$\delta$(Ni) & -0.13$\pm$0.14 & -0.33$\pm$0.05 & -0.33$\pm$0.10 & -1.46$\pm$0.04 & -1.04$\pm$0.02 & -0.80$\pm$0.04 & -2.79$\pm$0.09 & -1.75$\pm$0.03 & -1.26$\pm$0.03 \\
$\delta$(Zn) & -0.33$\pm$0.05 & 0.00$\pm$0.06 & 0.00$\pm$0.12 & -0.48$\pm$0.04 & -0.23$\pm$0.02 & -0.10$\pm$0.05 & -0.64$\pm$0.10 & -0.66$\pm$0.03 & -0.51$\pm$0.04 \\
\cline{1-10}
D$/$M & 0.30$\pm$0.08 & 0.12$\pm$0.11 & 0.08$\pm$0.14 & 0.43$\pm$0.06 & 0.34$\pm$0.07 & 0.16$\pm$0.10 & 0.52$\pm$0.04 & 0.52$\pm$0.05 & 0.40$\pm$0.06 \\
%D\/M (w\/o C, O) & 0.69$\pm$0.02 & 0.33$\pm$0.06 & 0.23$\pm$0.13 & 0.89$\pm$0.02 & 0.79$\pm$0.01 & 0.50$\pm$0.04 & 0.96$\pm$0.01 & 0.93$\pm$0.01 & 0.86$\pm$0.01 \\
\enddata
\tablecomments{For C and O, we estimate depletions by applying the MW relation between Fe and C or O depletion (see Section \ref{estimating_c_and_o})}
\tablecomments{Depletions are theoretically capped at 0, resulting in some 0.00 values in the fits of depletions vs $\log$ N(H), particularly at low column density and for volatile elements}
\end{deluxetable*}

\subsection{Relation between D/G and hydrogen column density}\label{section_dg_nh}

\begin{figure*}
\centering
\includegraphics[width=\textwidth]{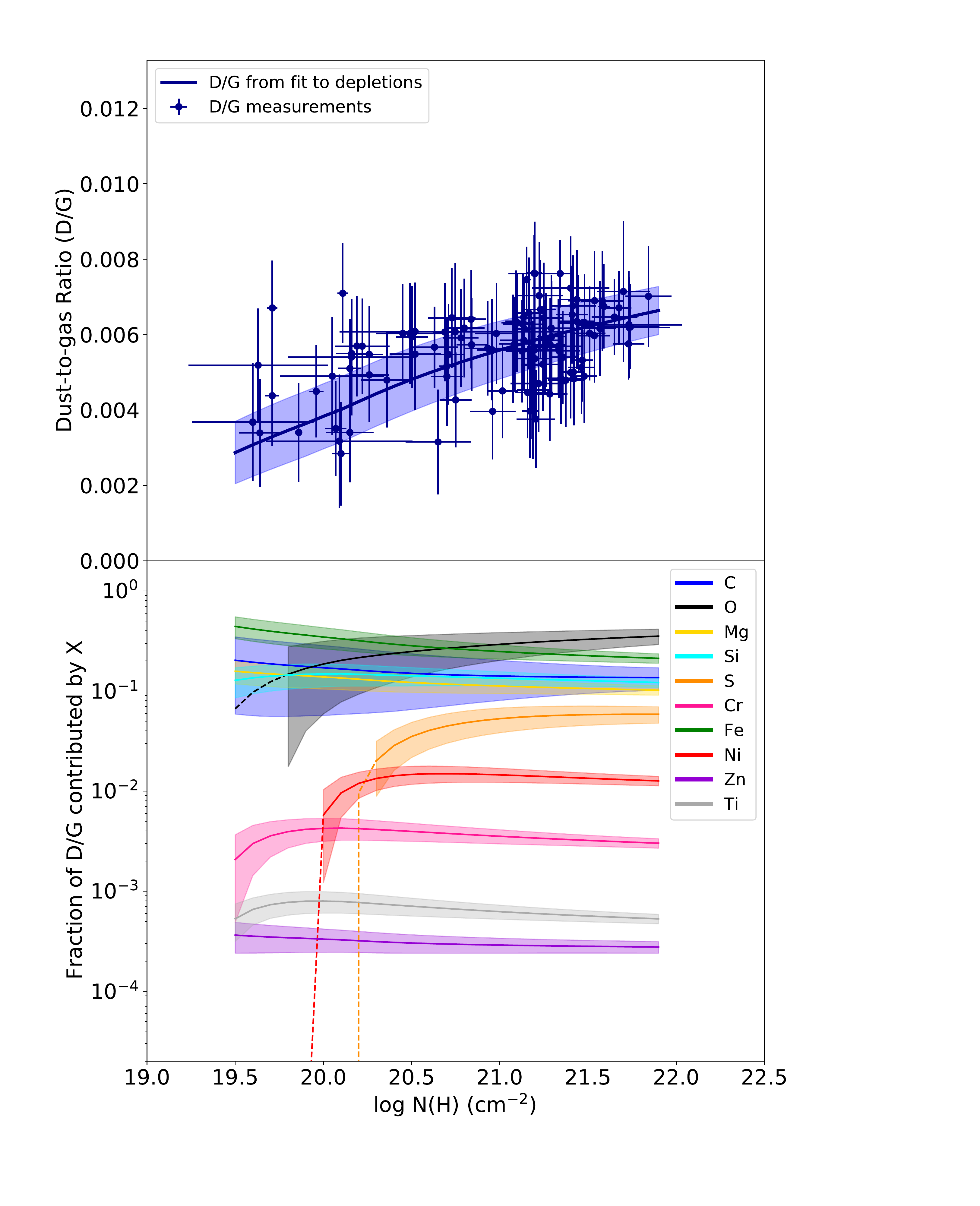}
\caption{(Top) Dust-to-gas ratio (D/G) in the MW, obtained from the collection of depletions measured by \citet{jenkins2009}, as a function of the logarithm of the hydrogen column density, N(H) (blue points and band). The points are measurements for each sight-line, while the blue line and band were obtained from the fits of the individual depletions with $\log$ N(H) and their 1$\sigma$ uncertainty (coefficient of the fits are given in Table 3). (Bottom) Fraction of the dust mass contributed by each element, as a function of $\log$ N(H). The fraction of D/G contributed by some elements (e.g., S, Ni) drops to zero at low column densities due to depletions crossing the theoretical maximum limit of zero.}
\label{plot_dg_mw}
\end{figure*}

\begin{figure*}
\centering
\includegraphics[width=\textwidth]{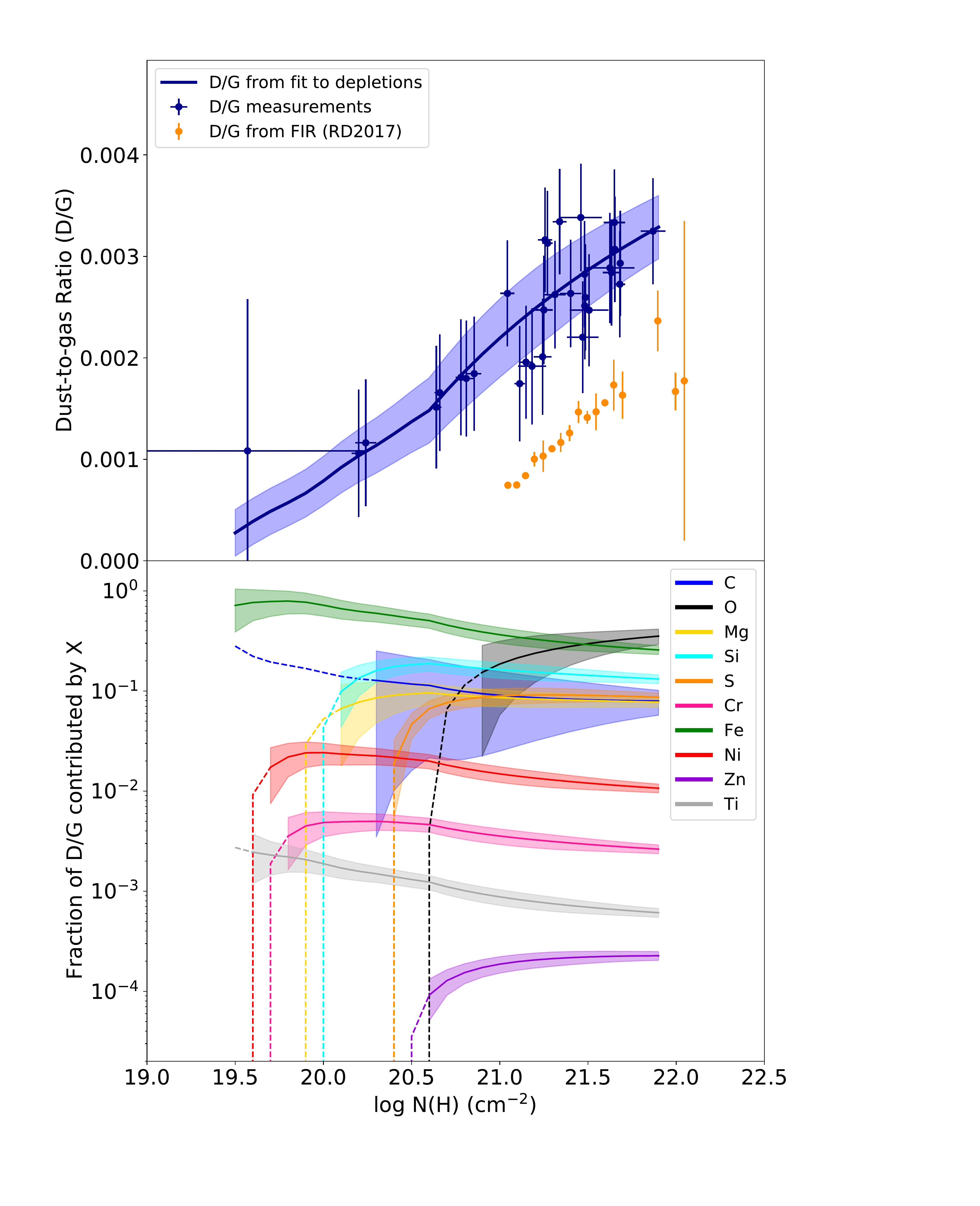}
\caption{(Top) Dust-to-gas ratio (D/G) in the LMC, obtained from the collection of depletions measured by the METAL program \citep{RD2021}, as a function of the logarithm of the hydrogen column density, N(H) (blue points and band). The points are measurements for each sight-line, while the blue line and band were obtained from the fits of the individual depletions with $\log$ N(H) and their 1$\sigma$ uncertainty (coefficient of the fits are given in Table 3). For comparison, the D/G measured from FIR, 21 cm, and CO (1-0) emission in \citet{RD2017} is shown in black. (Bottom) Fraction of the dust mass contributed by each element, as a function of $\log$ N(H). The fraction of D/G contributed by some elements (e.g., O, Mg, Si, S) drops to zero at low column densities due to depletions crossing the theoretical maximum limit of zero.}
\label{plot_dg_lmc}
\end{figure*}

\begin{figure*}
\centering
\includegraphics[width=\textwidth]{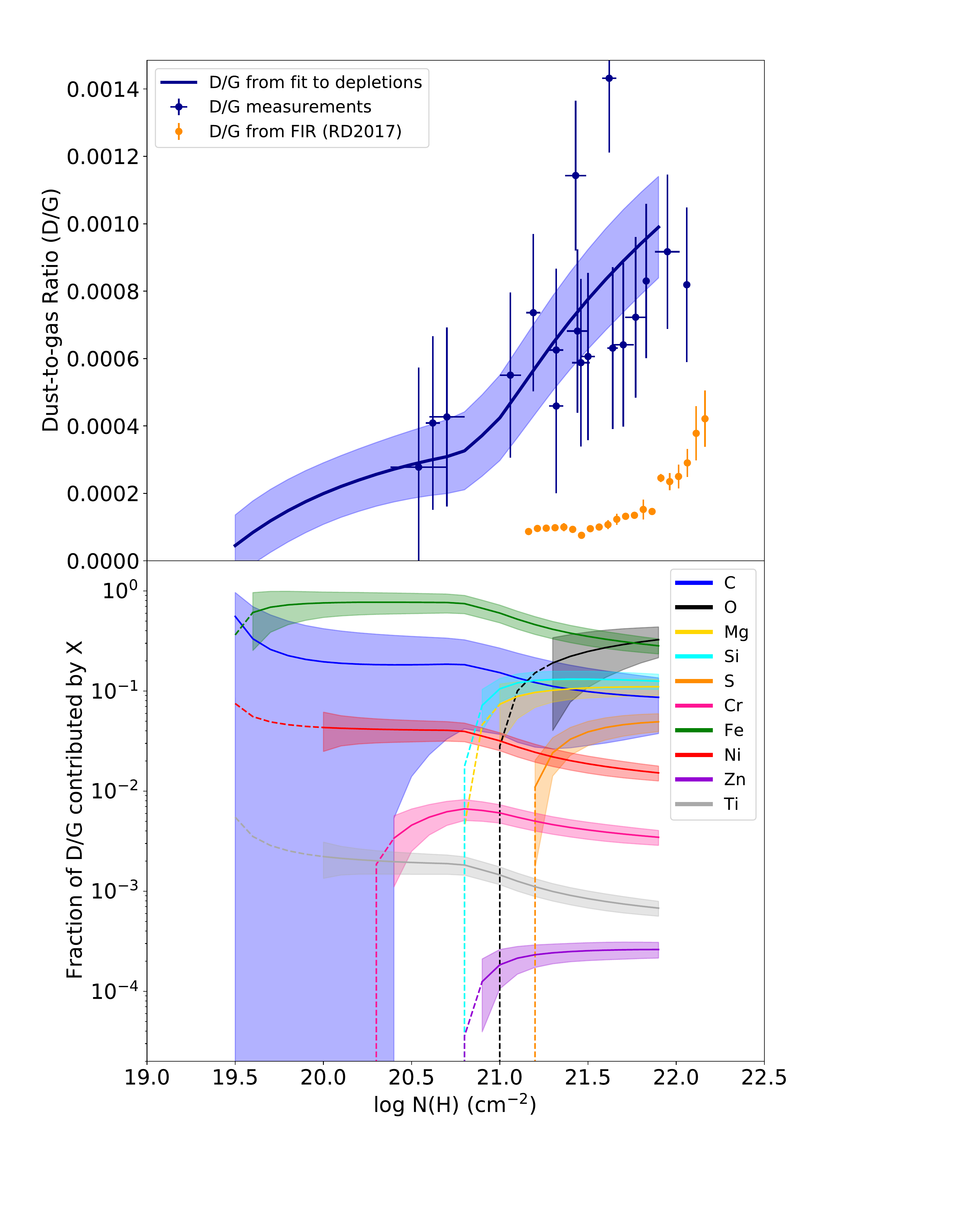}
\caption{(Top) Dust-to-gas ratio (D/G) in the SMC, obtained from the collection of depletions measured by \citet{jenkins2017}, as a function of the logarithm of the hydrogen column density, N(H) (blue points and band). The points are measurements for each sight-line, while the blue line and band were obtained from the fits of the individual depletions with $\log$ N(H) and their 1$\sigma$ uncertainty (coefficient of the fits are given in Table 3). For comparison, the D/G measured from FIR, 21 cm, and CO 1-0 in \citet{RD2017} is shown in black. (Bottom) Fraction of the dust mass contributed by each element, as a function of $\log$ N(H). The fraction of D/G contributed by some elements (e.g., O, Mg, Si, S) drops to zero at low column densities due to depletions crossing the theoretical maximum limit of zero.}
\label{plot_dg_smc}
\end{figure*}

\begin{figure}
\centering
\includegraphics[width=8cm]{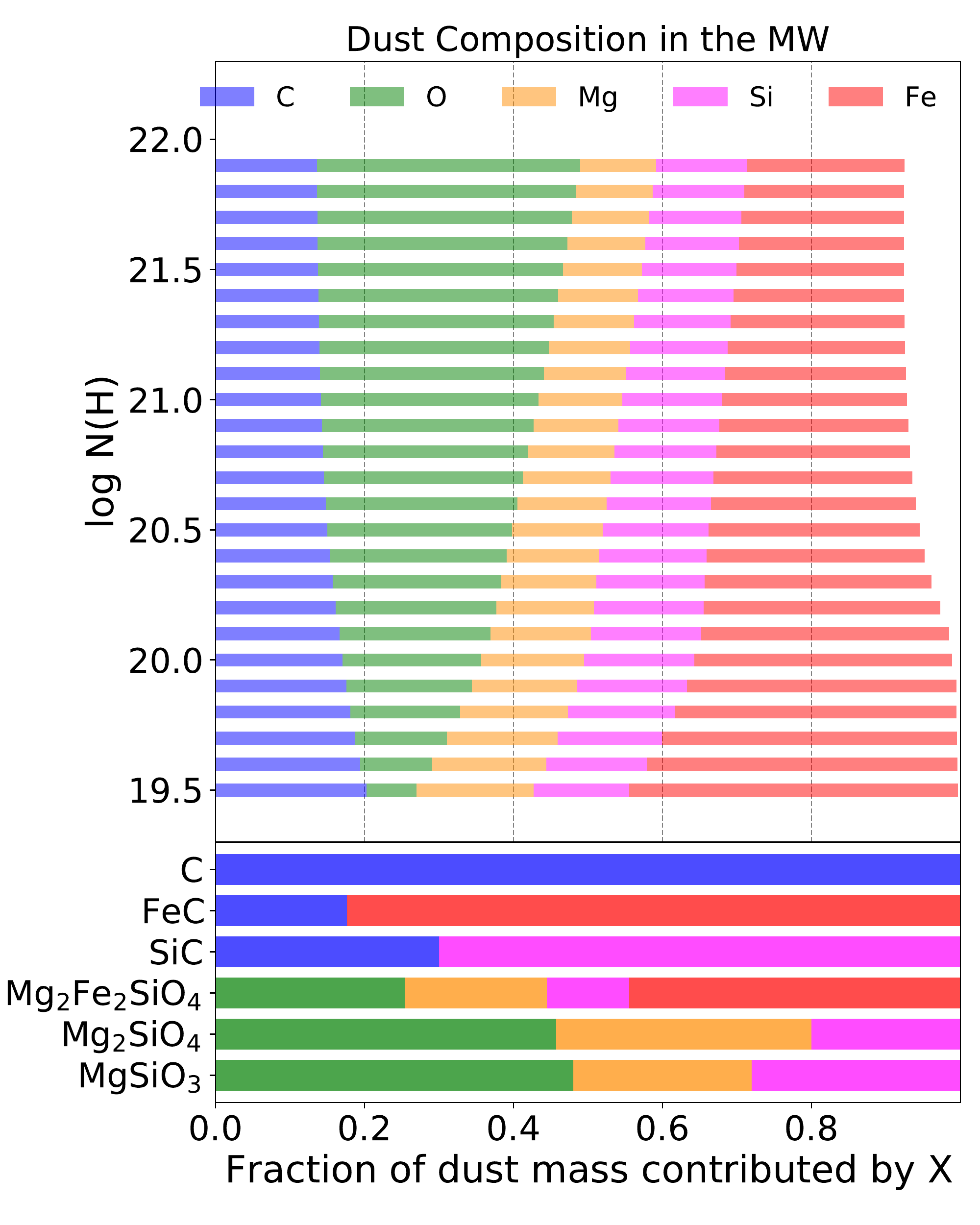}
\caption{(Top) Dust composition (i.e., fraction of the dust mass contributed by each element C, O, Mg, Si, Fe) as a function of $\log$ N(H) in the MW. At high column densities, the fractions of the dust mass from C, O, Mg, Si, and Fe do not add up to one due to the contribution from other elements (S, Ni) not plotted here. (Bottom) Mass fraction of C, O, Mg, Si, Fe for known condensates: C (graphite), FeC (iron carbide), SiC (silicon carbide), and silicates (olivine (Mg$_2$Fe$_2$SiO$_4$), forsterite (Mg$_2$SiO$_4$), enstatite (MgSiO$_3$)).   }
\label{plot_dust_composition_MW}
\end{figure}

\begin{figure*}
\centering
\includegraphics[width=8cm]{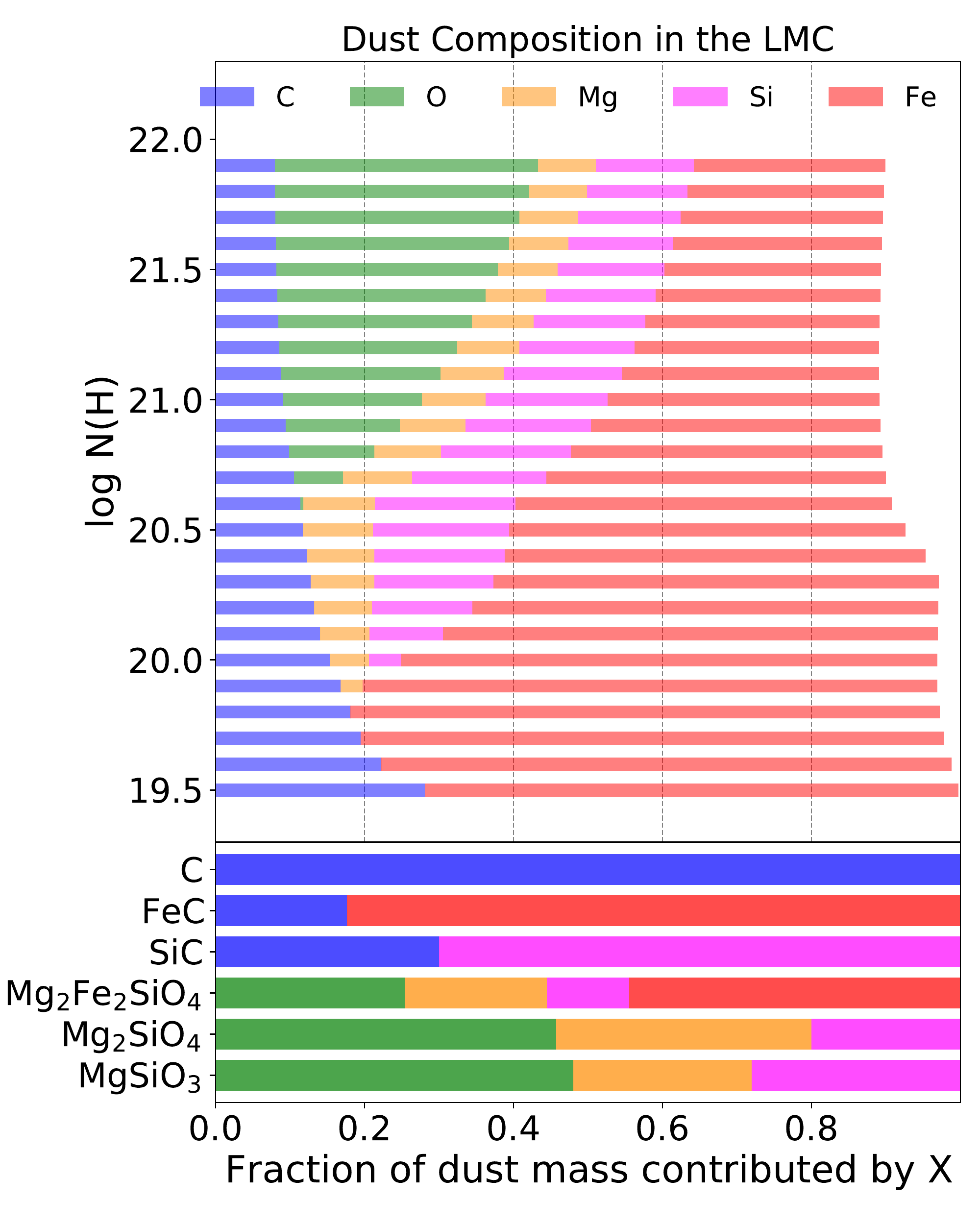}
\includegraphics[width=8cm]{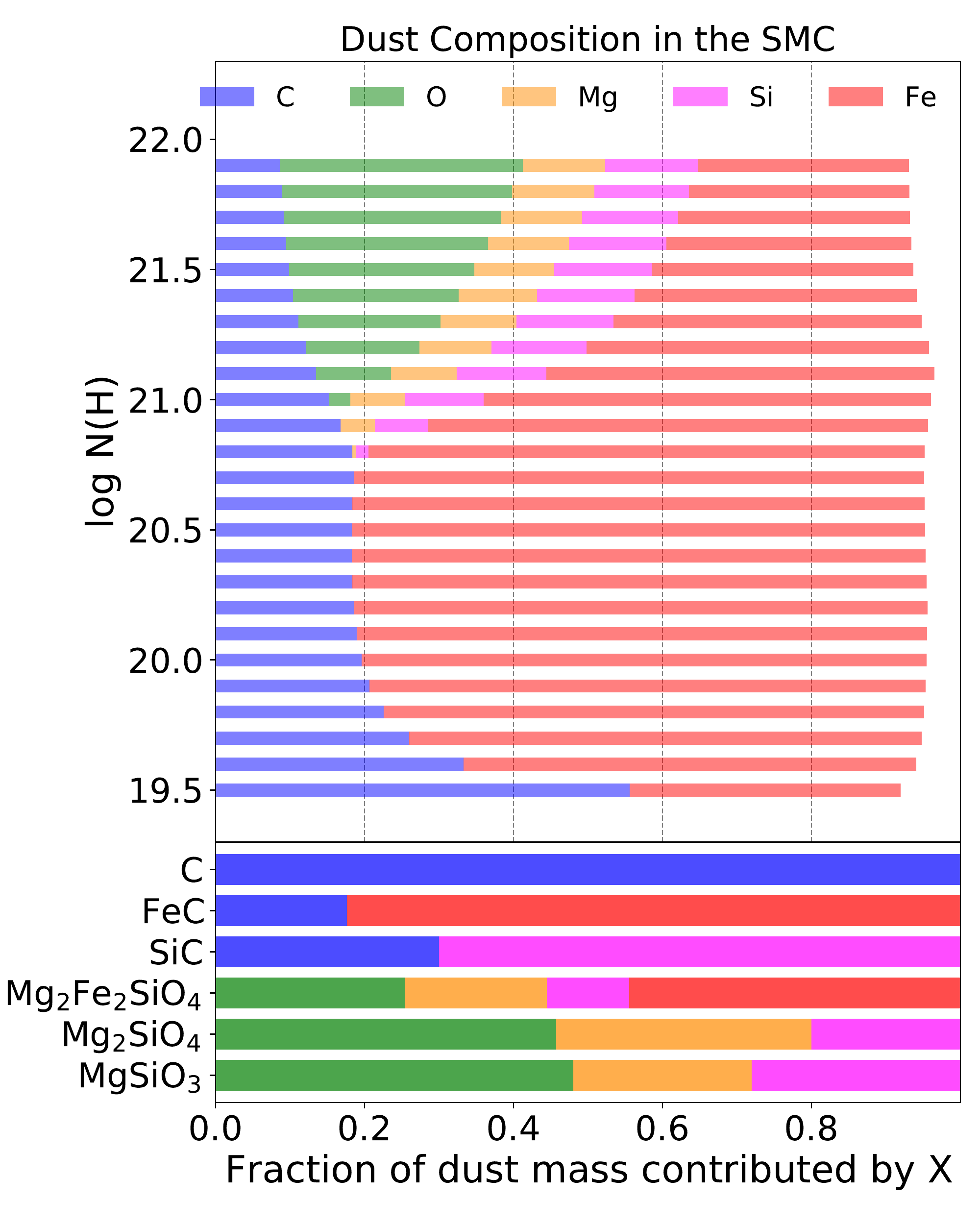}
\caption{Same as Figure \ref{plot_dust_composition_MW}, but for the LMC (left) and SMC (right).}
\label{plot_dust_composition_LMC_SMC}
\end{figure*}

\indent Applying Equation \ref{dg_eq} to the depletion measurements of Mg, Si, S, Cr, Fe, Ni, Zn, Ti and the estimation of depletions for C and O, we obtain the relation between D/G and column density shown in the top panels of Figures \ref{plot_dg_mw}, \ref{plot_dg_lmc} and \ref{plot_dg_smc} for the MW, LMC, and SMC. In those figures, we plot both the D/G computed for individual sight-lines (red points), and the relation between D/G and $\log$ N(H) derived from the fits of depletions to $\log$ N(H) given in Table 3 (blue lines). Errors on the $A_{\mathrm{H}}$(X) and $B_{\mathrm{H}}$(X) coefficients are included in the calculation of the fitted relation between D/G and $\log$ N(H).\\
\indent In all three galaxies and consistent with the trends we observe for depletions of individual elements, the depletion-based D/G increases with increasing hydrogen column density. Between $\log$ N(H) $=$ 20 and 22 cm$^{-2}$, the increase in D/G represents a factor 2 in the MW, 4 in the LMC, and 5 in the SMC. \\
\indent The observed trend of D/G versus $\log$ N(H) is consistent with the theoretical expectation that gas-phase metals accrete onto dust grains at higher rates in higher density environments, as explained in \citet{asano2013} for example: the timescale for dust growth in the iSM is inversely proportional to density. The correlation between depletions (or D/G) vs $\log$ N(H) has more scatter in the MW than in the LMC and SMC, presumably due to the effects of varying path lengths in the MW. Indeed, sight-lines in the MW go through a longer and varying path in the disk, while sight-lines in the LMC and SMC probe gas and dust in their disks face-on (to a lesser extent in the SMC owing to its "cigar" shape, although \citet{Yanchulova-Merica-Jones2021} demonstrate that gas in the SMC constitutes a thin layer). In the LMC, and to a lesser extent the SMC, variations in the path length are driven by changes in the scale height of the gas perpendicular to the plane of the LMC or SMC. The magnitudes of such variations are probably small compared to the variation of N(H) in the ISM of these galaxies. Hence, $\log$ N(H) should be a good proxy for the average n(H) over the entire line of sight to a star embedded near the plane of the LMC or SMC. On the other hand, the path length of a line-of-sight in the MW can vary considerably depending on the distance to the background star, and $\log$ N(H) does not trace the mean density along the line of sight, resulting in more scatter in the relation between depletions and $\log$ N(H). Correspondingly, in the case of the MW, where the path length through the ISM $d$ can be determined, \citet{jenkins2009} showed that depletions (through $F_*$) correlate much better with N(H)/$d$ than with N(H). \\
\indent In the LMC and SMC, D/G is also measured from FIR emission in \citet{RD2014, RD2017}, particularly as a function of $\log$ N(H) in the latter study, and we plot these trends in Figures \ref{plot_dg_lmc} and \ref{plot_dg_smc} for comparison with the depletion-based D/G measurements. This comparison for the LMC was discussed in \citet{RD2021}: the FIR-based D/G is a factor 2 lower than the depletion-based D/G, but the slopes of D/G vs $\log$ N(H) are similar for both types of measurements. In the SMC, the FIR-based D/G is a factor 3 lower than the depletion-based D/G for $\log$ N(H) $>$ 20.5 cm$^{-2}$, but the discrepancy is larger (factor of $\sim$5) at lower column densities. The possible explanations for this discrepancy are presented in Section \ref{fir_tension}.

\section{The dust composition inferred from depletions}\label{composition_section}
\indent Depletion measurements provide some clue as to the composition of dust in the MW, LMC and SMC. Indeed, the fraction of the dust mass contributed by each element X, $D_{\mathrm{X}}$, is given by:

\begin{equation}
D_{\mathrm{X}} = \frac{(1-10^{\delta(\mathrm{X})})  \left (\frac{N(X)}{N_\mathrm{H}} \right )_{\mathrm{tot}} W(\mathrm{X})}{1.36 (D/G)}
\end{equation}

%(\frac{N(X)}{N_{\mathrm{H}} \right )_{\mathrm{tot}} 

\noindent where $\delta$(X) are the depletions, and $A$(X) and $W$(X) are the same terms as in Equation \ref{doh_equation}. $D_{\mathrm{X}}$ for the the elements measured in the samples is shown in the bottom panels of Figures \ref{plot_dg_mw}, \ref{plot_dg_lmc} and \ref{plot_dg_smc} for the MW, LMC, and SMC, respectively. Not surprisingly, the dust mass is dominated by C, O, Mg, Si, Fe, and to a lesser extent at the highest column densities, S.  In the MW, the dust mass budget is roughly equally split between C, O, Mg, Si, Fe. However, as the metallicity decreases from the MW, to the LMC and SMC, the contribution of Mg, Si, O kicks in at increasingly higher hydrogen column densities, roughly $\log$ N(H) $\sim$ 20.5 cm$^{-2}$ in the LMC and 21 cm$^{-2}$ in the SMC. \\
\indent To visualize the dust composition in the MW, LMC and SMC more effectively, bar plots of the dust mass fraction contributed by C, O, Mg, Si, and Fe are shown as a function of $\log$ N(H) in each galaxy in Figures \ref{plot_dust_composition_MW} and \ref{plot_dust_composition_LMC_SMC}. Additionally, the fraction of the dust mass contributed by these elements in known condensates such as olivine, enstatite, or iron carbide are shown for comparison. No single condensate matches the observed composition of dust in the MW, LMC, or SMC, indicating a mix of different dust types is present in those galaxies. However, carbonaceous grains and iron carbide must dominate the dust mass at low column densities in the LMC (below $\log$ N(H) = 20.5 cm$^{-2}$) and SMC (below $\log$ N(H) = 21 cm$^{-2}$), while the contribution from silicates increases with increasing column density in those galaxies. This would be in line with the dust properties observed in the LMC and SMC using the FIR \citep{chastenet2017}, where carbon dust is observed to dominate as evidence by the spectral emissivity index of the FIR SED.

\section{Depletions, D/M, and D/G versus metallicity}\label{section_dg_vs_Z}

\indent Chemical evolution models, such as \citet{asano2013} or \citet{feldmann2015} predict that metallicity is an important factor in setting the abundance of dust in galaxies (in addition to density, as seen in Section \ref{section_nh}). Fundamentally, this is expected because the timescale for dust growth in the ISM is inversely proportional to metallicity \citep[see, e.g., Equation 20 in][]{asano2013}. With large samples of depletions in the MW, LMC ($Z$ $=$ 0.5\Zsun), and SMC (Z $=$ 0.2 \Zsun), we can put constraints on the variations of depletions, D/M, and D/G with metallicity down to 20\% solar metallicity. 

\subsection{Depletions and D/M versus metallcity} \label{deps_dm_vs_metallicity}

\begin{figure*}
\centering
\includegraphics[width=\textwidth]{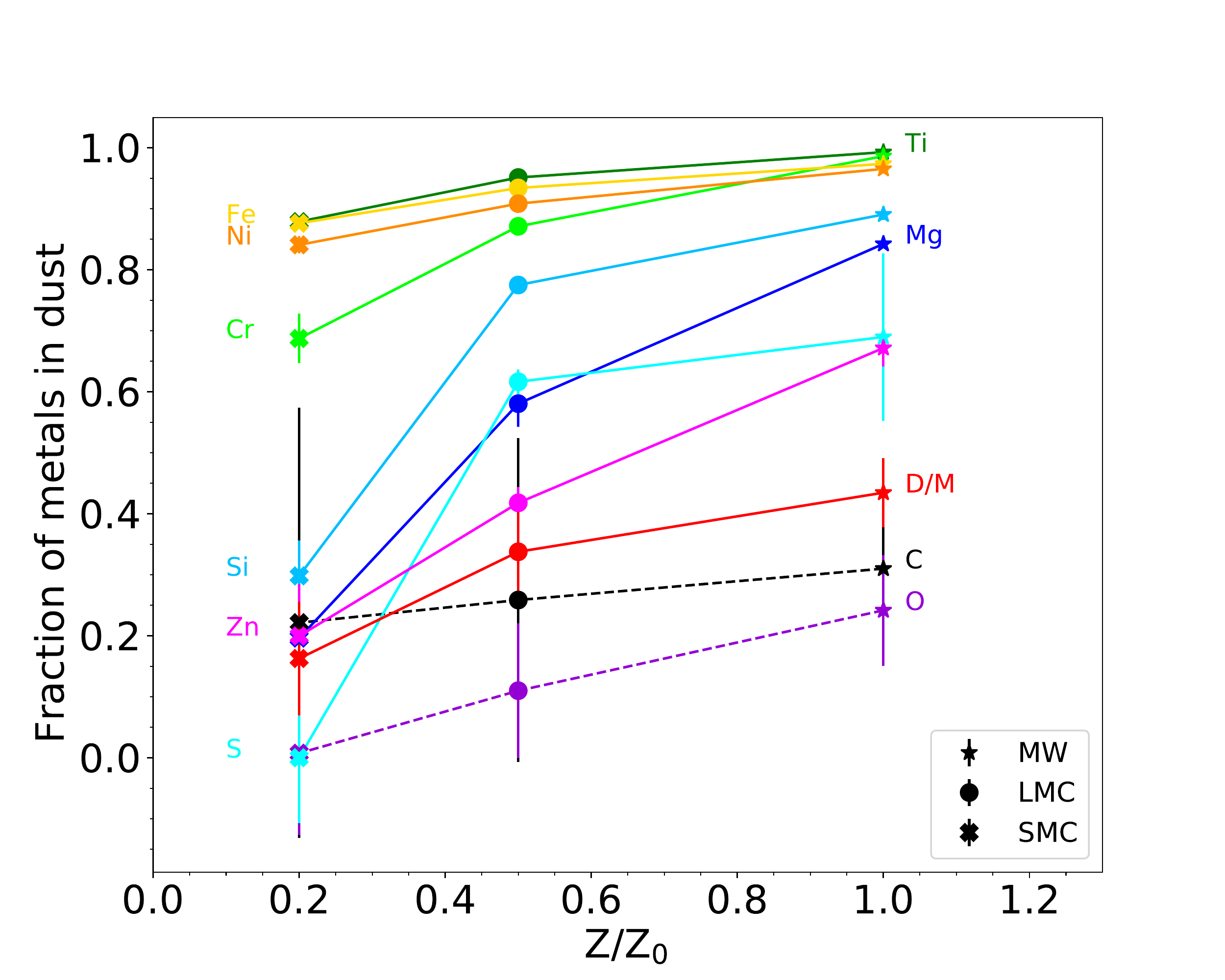}
\caption{Fraction of elements C, O, Mg, Si, S, Cr, Fe, Ni, Zn, Ti in the dust as a function of metallicity based on measurements the MW, LMC, and SMC at $\log$ N(H) $=$ 21 cm$^{-2}$. The D/M, obtained from the mass weighted sum of abundances of elements in the dust, is also plotted (also for $\log$ N(H) $=$ 21 cm$^{-2}$. }
\label{plot_dust_fractions}
\end{figure*}

\indent Despite the relatively small differences in the slopes of the $\log$ N(H)---$\delta$(X) relation between the MW, LMC, and SMC seen in Figure \ref{plot_deps_nh}, a clear trend with metallicity emerges: for all elements, the relation between $\log$ N(H) and $\delta$(X) lies lowest for the MW (most depleted), highest for the SMC (least depleted), with the LMC between those two extremes. This trend was already observed in \citet{RD2019} based on Si only. Taking the mean depletion difference between the MW and LMC or SMC from Table 4 for all elements we can measure in all three galaxies (Mg, Si, S, Cr, Fe, Ni, Zn, Ti), and weighting the mean depletion difference by the inverse square of the errors, we obtain mean weighted depletion differences of 0.46$\pm$0.01 between the MW and LMC, and 0.80$\pm$0.02 dex between the MW and SMC, at $\log$ N(H) $=$ 21 cm$^{-2}$. This implies that the fraction of metals other than C and O in the gas is 2.9 times higher in the LMC than in the MW, and 6.3 times higher in the SMC than the MW. \\
\indent Because the fraction of metals locked in dust, given by 1-10$^{\delta(X)}$, remains fairly high, even at the metallicity of the SMC, the differences in D/M between the three galaxies are much less pronounced that the differences in gas-phase fractions, given by 10$^{\delta(X)}$. This is shown in Figure \ref{plot_dust_fractions}, where we plot the fraction of elements in dust and the D/M as a function of metallicity (using the depletion values in Table 4) for $\log$ N(H) $=$ 21 cm$^{-2}$. For example, the fraction of Fe in the gas-phase is about 5 times higher in the SMC than in the MW (0.67 dex difference in depletion at $\log$ N(H) $=$ 21 cm$^{-2}$, see Table 4), but the fraction of Fe in the dust phase only subsequently decreases from 97\% (MW) to 88\% (SMC). The fraction of C and O in the dust is relatively low, which could in principle lead to larger variations in dust-phase fractions between the MW, LMC, and SMC. However, the differences in gas-phase C and O fractions between the MW, LMC and SMC are not very large. The reason for this small difference is that the slope $A_{\mathrm{C}}$ and $A_{\mathrm{O}}$ of depletions versus $F_*$ in the MW are very shallow ($-$0.101 and $-$0.22 respectively). Thus, even if the depletions of Fe (and all other elements except C and O) in the LMC and SMC are 0.4 dex and 0.7 dex less negative than in the MW, corresponding to a lower $F_*$ by $-$0.25 and $-$0.5, respectively, the corresponding difference in C depletions is only 10\% of the difference in $F_*$, hence 0.03 dex between the MW and LMC, and 0.05 dex between the MW and the SMC. Similarly, the difference in O depletions is 22\% of the difference in $F_*$, or 0.06 dex between the MW and LMC, and 0.11 dex between the MW and SMC. As a result, the D/M in the LMC is only 25\% lower than in the MW at $\log$ N(H) $=$ 21 cm$^{-2}$. In the SMC, the D/M is a factor 3 lower than in the MW at $\log$ N(H) $=$ 21 cm$^{-2}$.

\subsection{Impact of the varying D/M on neutral gas-phase metallicities in the MW, LMC, and SMC}

\begin{figure*}
\centering
\includegraphics[width=\textwidth]{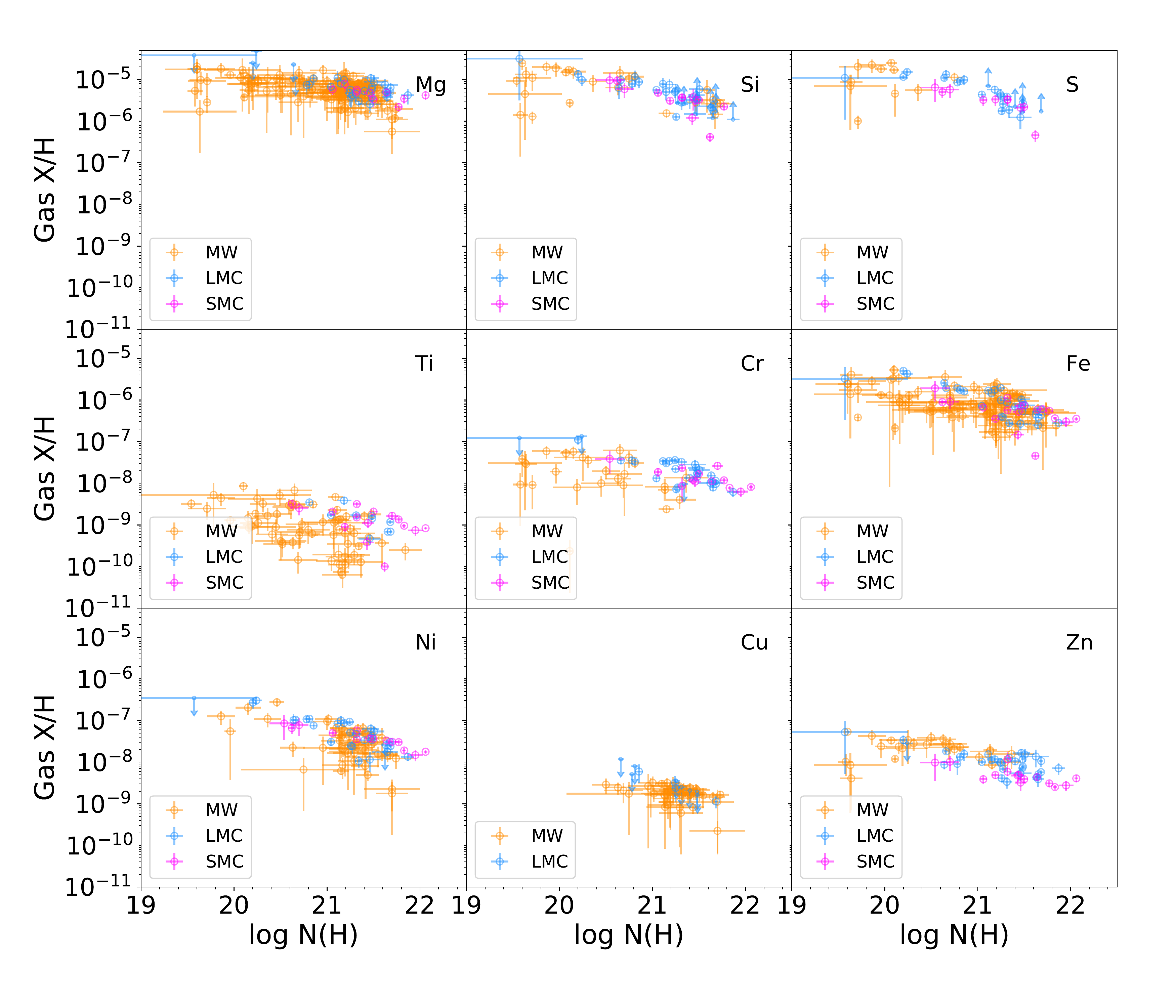}
\caption{Gas-phase abundances of Mg, Si, S, Cr, Fe, Ni, Cu, Zn, Ti, Cr, as a function of $\log$ N(H). Orange, blue, and magenta correspond to the Milky Way \citep{jenkins2009}, LMC \citep{RD2021}, and SMC \citep{jenkins2017}, respectively. For a given hydrogen column density, the gas-phase abundances in the MW, LMC, and SMC are similar, despite the total metallicity of these galaxies differing by factors of 2 and 5 respectively. This is due to the lower fraction of metals locked into dust grains at low metallicity.}
\label{plot_gasmet_nh}
\end{figure*}

\indent The trends of depletions and D/M with metallicity observed in Section \ref{deps_dm_vs_metallicity} have a surprising result on the neutral gas-phase metallicities of the MW, LMC, and SMC. Because the total metallicity of the MW is 5 times higher than that of the SMC, but the fraction of metals (other than C and O) in the gas in the MW is 6 lower than in the SMC, the neutral gas-phase metallicities of these two galaxies should be about the same for a given hydrogen column density. Similarly, the total metallicity of the LMC is twice lower than that of the MW, but the fraction of metals in the gas in the LMC is 2.5 times higher than in the MW, and so the neutral gas-phase metallicities of the MW and LMC should be similar at a fixed hydrogen column density. Figure \ref{plot_gasmet_nh} confirms that, for a given $\log$ N(H), the neutral gas-phase metallicities of the MW, LMC, and SMC are about the same, despite the masses of these galaxies differing by two orders of magnitude (M$_*$(MW) $\sim$ 6$\times 10^{10}$ \Msu \citep{licquia2015}; M$_*$(LMC) $=$ 2.7$\times 10^9$ \Msu \citep{vandermarel2006}; and M$_*$(SMC) $=$ 3.1$\times 10^8$ \Msu \citep{besla2015}). \\
\indent A direct consequence of this effect is that, without further information on the depletion levels from abundance ratios \citep[e.g.,][]{decia2016, decia2018a}, DLA systems with metallicities similar to the MW and Magellanic Clouds at high redshift would be indistinguishable based on their gas-phase metallicities measured from QSO spectroscopy. This is particularly important given that volatile elements such as S and Zn used as metallicity tracers in such systems, do deplete from the gas-phase. Recovering the total metallicity of volatile elements in DLAs therefore requires accurate depletion corrections. Such corrections have been derived using abundance ratios in DLAs ([Zn/Fe]) and the MW calibration of the relation between $\delta$(Zn) and [Zn/Fe] \citep{decia2016}. This effect provides additional motivation for deriving calibrations of the [Zn/Fe]---depletion relation in the MW, LMC, and SMC, testing them on DLAs, which will be presented In the upcoming METAL IV paper (Roman-Duval et al., in prep).

\subsection{D/G versus metallicity from depletions} 

\begin{deluxetable*}{cc|cc|cc}
\tablenum{5}
\tablecaption{D/G values obtained from depletions and FIR in the MW, LMC, and SMC}
\tablewidth{0pt}
\tablehead{
\colhead{} &  \colhead{MW} & \multicolumn{2}{c}{LMC} & \multicolumn{2}{c}{SMC}\\
& Depletions &  Depletions & FIR\tablenotemark{a}  & Depletions & FIR\tablenotemark{a} \\
}
\startdata
D/G $\log$ N(H) $=$ 20 cm$^{-2}$   &  (3.83$\pm$1.03)$\times10^{-3}$       &   (7.87$\pm$7.6)$\times10^{-4}$     &  (7.43$\pm$0.58)$\times10^{-4}$  &     (1.99$\pm$3.51)$\times10^{-4}$  & (8.68$\pm$1.91)$\times10^{-5}$  \\
D/G $\log$ N(H) $=$ 21 cm$^{-2}$   &   (5.59$\pm$0.74)$\times10^{-3}$    &    (2.19$\pm$0.49)$\times10^{-3}$    &  (8.38$\pm$0.66)$\times10^{-4}$   &    (4.24$\pm$2.56)$\times10^{-4}$    &    (8.68$\pm$1.91)$\times10^{-5}$ \\
D/G $\log$ N(H) $=$ 22 cm$^{-2}$   &   (6.74$\pm$0.57)$\times10^{-3}$    &    (3.38$\pm$0.30)$\times 10^{-3}$   &  (1.77$\pm$3.15)$\times10^{-3}$   &    (1.03$\pm$0.16)$\times10^{-3}$      &    (3.78$\pm$1.60)$\times10^{-4}$\\
D/G Integrated                                  &    (5.98$\pm$0.65)$\times10^{-3}$    &    (2.30$\pm$0.11)$\times10^{-3}$   &   (1.27$\pm$0.12)$\times10^{-3}$ &    (7.57$\pm$0.8)$\times10^{-4}$    &    (1.56$\pm$0.17)$\times 10^{-4}$ \\
\enddata
\tablenotetext{a}{The FIR D/G values are from \citet{RD2017}}
\end{deluxetable*}

\begin{figure*}
\centering
\includegraphics[width=\textwidth]{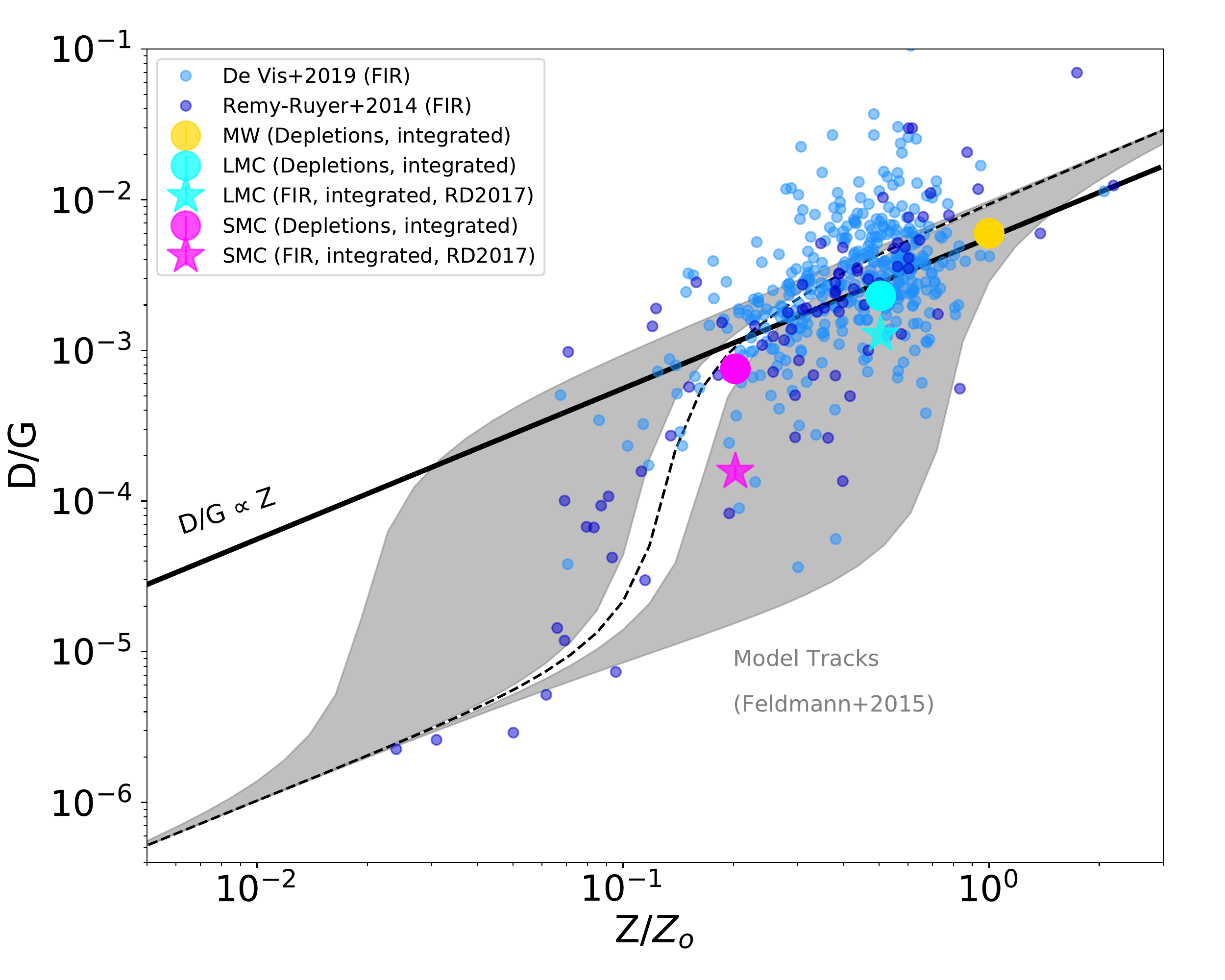}
\caption{Dust-to-gas ratio as a function of total (gas + dust) metallicity in different systems and from different observational methods. The blue points correspond to the D/G measured in nearby galaxies using FIR emission to trace dust, and 21 cm and CO rotational emission to trace atomic and molecular gas \citep{remy-ruyer2014, devis2019}. The cyan and magenta stars correspond to similar measurements in the LMC and SMC, respectively \citep{RD2017}. The FIR measurements are integrated (total dust mass/total gas mass). The yellow, cyan and magenta circles show D/G measurements obtained from spectroscopic depletions in the MW, LMC, and SMC \citep{jenkins2009, RD2021, jenkins2017}. The D/G estimated from depletions is integrated using the approach described in Section \ref{fir_tension}. Lastly, the gray tracks show the chemical evolution model from \citet{feldmann2015} for a range of the $\gamma$ parameter (2$\times 10^3$ --- $10^6$).}
\label{plot_feldmann}
\end{figure*}

\indent We estimate the D/G in each galaxy (MW, LMC, SMC) for a given $\log$ N(H) using the approach described in Section \ref{section_dg_nh}, by basically summing the mass weighted dust fractions of the elements for which we measure depletions, as well as C and O for which we estimate depletions using the method outlined in Section \ref{estimating_c_and_o} (see also Table 4 for numerical values of depletions used in the calculation of D/G). The resulting D/G values for $\log$ N(H) $=$ 20, 21, and 22 cm$^{-2}$ are listed in Table 5. The D/G in the MW, LMC, and SMC is plotted in Figure \ref{plot_feldmann} for $\log$ N(H) = 21 cm$^{-2}$. At $\log$ N(H) $=$ 21 cm$^{-2}$, the D/G in the LMC is a factor 2.6 lower than in the MW, while the SMC D/G is 13 times smaller than in the MW. The latter value of D/G in the SMC is in excellent agreement with result of \citet{Yanchulova-Merica-Jones2021}, who found from extinction modeling that $<A_V/N_H>$ in the SMC is 14 times smaller than in the MW. \\
\indent This decrease of D/G with metallicity is steeper than linear, as expected from a D/M that also decreases with metallicity. Indeed, we showed in Section \ref{deps_dm_vs_metallicity} that the SMC D/M is a factor 2---3 lower than in the MW. The variation of D/G with metallicity inferred from depletion measurements in the MW, LMC, and SMC are consistent with the chemical evolution model from \citet{feldmann2015} (plotted in Figure \ref{plot_feldmann}) that takes into account dust formation in evolved stars, dust growth in the ISM, dust destruction by SNe shocks, and dust dilution by inflows of pristine gas. The model is plotted for a plausible range of the parameter $\gamma$, which is the ratio of the molecular gas consumption by star-formation timescale \citep[typically 2 Gyr, see][]{bigiel2008} to the timescale for dust growth in the ISM in the MW \citep[typically 10 Myr, see ][]{hirashita2000, asano2013, feldmann2015}. In Figure \ref{plot_feldmann}, $\gamma$ ranges from 2$\times 10^3$ to $10^6$ with a fiducial value $\gamma$ $=$ 3$\times 10^4$. In the model, above a critical metallicity at which the dust input rate from evolved stars (AGB + supernovae) and ISM dust growth balances the dust destruction by supernova (SN) shockwaves and dilution by inflows of pristine gas, the D/M is high with most metals locked in the dust-phase. Below this critical metallicity, the D/M is low and determined by the input of stellar dust sources. The metallicity of the SMC (20\% solar) still lies above the critical metallicity where the D/M and therefore D/G starts to decrease steeply with metallicity. Depletion measurements at metallicities 10\% solar or lower are needed to fill this gap in our understanding of the dust abundance and chemical evolution of galaxies. This is the subject of an ongoing investigation using data taken as part of the HST large program METAL-Z (G0-15880, Hamanowicz et al., in prep). \\

\subsection{D/G versus metallicity: comparison with FIR measurements}\label{fir_tension}

\indent The D/G estimates in the MW, LMC, and SMC derived from depletion measurements complement previous FIR measurements of D/G vs metallicity \citep{RD2017, remy-ruyer2014, devis2019} in nearby galaxies, including the LMC and SMC. The FIR-based dust masses (or surface densities for resolved galaxies) are estimated by modeling the FIR SED observed with facilities such as Herschel, Spitzer, or Planck with either modified black bodies \citep[e.g.,][in the LMC and SMC]{gordon2014, RD2017} or full dust models \citep[e.g.][]{galliano2011, remy-ruyer2014, remy-ruyer2015, chastenet2017, devis2019, aniano2020, chastenet2021}. The atomic and molecular gas masses (or gas surface density) are estimated from \his 21 cm and CO rotational emission. The LMC and SMC are highly resolved, and therefore the D/G can be measured as a function of $\log$ N(H) as in \citet{RD2017}, or in an integrated manner (total dust mass/total gas mass).\\
\indent The D/G obtained from depletions corresponds to pencil beam sight-lines of a given $\log$ N(H), from which we can infer the trend of D/G vs $\log$ N(H) via linear fits, as done in Section \ref{section_dg_nh}. However, D/G measurements obtained from the FIR in nearby galaxies are integrated and correspond to the ratio of the total dust mass to the total gas mass. To estimate an integrated D/G from depletions and be able to compare it to integrated FIR-based measurements, we apply the relation between D/G and $\log$ N(H) established from depletions shown in Figures \ref{plot_dg_mw}, \ref{plot_dg_lmc} and \ref{plot_dg_smc},  (D/G)$_{\mathrm{dep}}$, to the $\log$ N(H) distribution observed with 21 cm emission (N(H)$_{\mathrm{21cm}}$) in the maps from \citet[][, SMC]{stanimirovic1999} and \citet[][, LMC]{kim2003}:

\begin{equation}\label{dg_integration}
(D/G)_{\mathrm{int}} = \frac{\sum_{\mathrm{pix}} N(H)_{\mathrm{21cm}} (D/G)_{\mathrm{dep}}(N(H)_{\mathrm{21cm}})}{\sum_{\mathrm{pix}}N(H)_{\mathrm{21cm}}}
\end{equation}

\indent The different D/G estimates in the MW, LMC, SMC (FIR and depletions) and nearby galaxies (FIR only) are shown in Figure \ref{plot_feldmann} as a function of (total) metallicity. For the LMC and SMC, we plot the integrated D/G. The chemical evolution model from \citet{feldmann2015} is also shown. A few key points stand out in Figure \ref{plot_feldmann}.  \\
\indent The first key conclusion from Figure \ref{plot_feldmann} is that the FIR-based and depletion-based D/G in the LMC and SMC differ by factors of 2 and 5, respectively, reflecting the differences previously observed in the relation between D/G and $\log$ N(H). \citet{RD2021} discussed possible reasons for this observed discrepancy. First, because the FIR surface brightness observed by Herschel, Spitzer, and Planck is the product of the dust surface density and dust opacity, an estimate of the dust surface density is degenerate with the assumed opacity, which can vary and is not well constrained observationally or theoretically \citep{stepnik2003, kohler2012, RD2014, demyk2017}. As a result, the FIR-based dust mass estimates suffer from systematic uncertainties of a factor of a few, even when the dust temperature can be accurately constrained using multi-band photometry. Observational constraints on the dust FIR opacity obtained by comparing the FIR emission observed in Herschel to extinction maps obtained from HST imaging at a range of metallicities are needed to resolve this discrepancy. This problem is the focus of several HST imaging programs in the LMC and SMC (Scylla, a parallel program to ULLYSES, PI Claire Murray; METAL GO-14675 \citep{RD2019}; SMIDGE \citep{yanchulova2017, Yanchulova-Merica-Jones2021}), in IC 1613 at 15\% solar metallicity and Sextans A at 8\% solar metallicity (GO-16513, PI Roman-Duval), and in Leo-P at 3\% solar metallicity (GO-16222, PI Christopher Clark). We note that, in estimating dust masses from the FIR, there is also a potential bias (under-estimation of the dust mass) due to the integrated nature of the measurement of dust surface densities that can vary on small scales \citep{galliano2011}. \\
\indent  Gas masses estimated from 21 cm and CO emission are not immune from such substantial systematics either. The molecular gas mass estimates rely on an assumed CO-to-H$_2$ conversion factor \citep{bolatto2013}, which is also poorly constrained and degenerate with D/G measurements \citep{RD2014}. Another potential issue in estimating atomic gas masses from 21 cm emission is that masses are often estimated from integrated measurements associated with a region that is spatially more extended than the region detected in dust emission (either on the sky or along the line of sight) , leading to a possible over-estimation of the gas mass. Thus, the systematic uncertainty on D/G estimates based on emission tracers could very well amount to a factor of several, perhaps up to an order of magnitude, and the effects describe above would preferentially under-estimate the D/G. \\
\indent On the depletions' side, the main source of systematic uncertainty is the poorly constrained contribution of C and O to the D/G estimated from depletions. Indeed, the estimates of C and O depletions rely on the MW relation between depletions of different elements, which might not apply at low metallicity due to different abundance ratios and subsequent chemical affinities. Additionally, the trend of C depletions vs $F_*$ in the MW were only measured toward a few sight-lines and are therefore highly uncertain (as shown by the error on the fits shown in Figure 5 of \citet{jenkins2009}). Therefore, it is possible that the depletion levels of C and O (and D/G) may be over- or under-estimated. We will have to wait until a LUVOIR-like observatory with enough UV sensitivity to observe the weak C and O lines outside the MW and observationally constrain the depletions of C and O at low metallicity. In the meantime, the uncertain contribution of C and O to the dust mass budget should be captured in our error bars (which explains why there are more likely fits to the trend of D/G vs $\log$ N(H) at higher D/G values than the fiducial relation in Figures \ref{plot_dg_mw}, \ref{plot_dg_lmc}, and \ref{plot_dg_smc}).\\
\indent A last source of discrepancy between the FIR and depletion-based D/G presented in \citet{RD2021} is the different geometrical set-ups of depletion and FIR observations can lead to different outcomes in the trend of D/G vs $\log$ N(H), as demonstrated based on simulations in \citet{RD2021}. \\

\section{Conclusion}\label{conclusion}
\indent We compiled and compared gas-phase abundances and depletions measurements in the Milky Way, LMC (50\% solar metallicity), and SMC (20\% solar metallicity. \\
\indent The relation between the depletions of Fe and that of other elements is relatively invariant between the MW, LMC and SMC, with the only exceptions of Mg and Ti showing 3--4$\sigma$ differences between the MW, LMC and SMC. Correspondingly, the relation between difference abundance ratios of refractory to volatile elements follow the same invariance between the three galaxies examined here. This implies that the depletion of Fe, which is easy to measure thanks to the numerous UV transitions of this element with a wide range of oscillator strengths, combined with the calibration of the $\delta$(Fe)---$\delta$(X) relation established in the Milky Way and Magellanic Clouds, can be used to estimate the depletion of elements in systems (such more distant galaxies) where depletions for elements other than Fe are difficult to observe spectroscopically. Such calibrations will be derived in the next METAL paper (METAL IV).\\
\indent In the MW, LMC, and SMC, the depletions of all elements observed become more negative (i.e., less metals in the gas-phase, more metals in the dust-phase) as the column density of hydrogen increases. Over the column density range $\log$ N(H) $=$ 20---22 cm$^{-2}$, the fraction of metals in the gas-phase typically decreases by 0.3 dex (Zn) to 1.3 dex (e.g., Fe, Ni, Cr, S, Si), but can decrease by as much as 2 dex (for Ti and Cr in the MW). As a result, the dust-to-gas ratio D/G increases by a factor 3 to 4 from $\log$ N(H) $=$ 20 to 22 cm$^{-2}$ in all three galaxies. This is consistent with the shorter timescales for accretion of gas-phase metals onto dust grains as the density of the ISM increases.\\
\indent By comparing the depletions in the MW, LMC, and SMC, we establish that the fraction of metals in the gas-phase increases with decreasing metallicity. The difference in the fraction of metals in the gas-phase amounts to a factor 3 between the MW and LMC, and a factor 6 between the MW and the SMC. Correspondingly, the fraction of metals locked in dust (the dust-to-metal ratio, D/M) decreases with decreasing metallicity. The D/M is a factor 1.2 lower in the LMC compared to the MW, and a factor 2--3 lower in the SMC than the MW. The immediate albeit surprising consequence of this variable D/M is that the neutral gas-phase metallicities of the MW, LMC, and SMC, for a given hydrogen column density, are very similar, despite the total metallicities of the LMC and SMC being lower than the MW's by factors of 2 and 5 respectively. Indeed, in the SMC (resp. LMC), the total abundance of metals is 5 (resp. 2) times lower than in the MW, but 6 (resp. 3) times less metals are locked away in dust grains, leaving the gas-phase abundances about the same as in the MW.\\
\indent By summing the depletions over the elements for which depletions are measured, as well as C and O for which we estimate depletions based on the MW relation between Fe and C or O depletions, we obtain D/G values of (5.59$\pm$0.74)$\times10^{-3}$, (2.19$\pm$0.49)$\times10^{-3}$, and (4.24$\pm$2.56)$\times 10^{-4}$ for the MW, LMC, and SMC at $\log$ N(H) $=$ 21 cm$^{-2}$. Integrating over the \his distribution observed via 21 cm emission, the D/G for the MW, LMC, and SMC are (5.98$\pm$0.65)$\times10^{-3}$, (2.30$\pm$0.11)$\times 10^{-3}$ and (7.57$\pm$0.80)$\times 10^{-4}$, respectively. \\
\indent We infer the dust composition in each galaxy from the depletions, and find that, while iron, carbon, and silicate components (Mg, Si, O) equally contribute to the dust budget in the MW, this is not the case in the LMC and SMC, where iron and carbon dominate the dust mass budget at all but the highest column densities ($\log$ N(H) $>$ 22 cm$^{-2}$).\\
\indent Since D/G $\propto$ D/M$\times$Z and D/M decreases with decreasing metallicity, the D/G observed through depletions decreases non-linearly (steeper than linearly) with metallicity. This is consistent with the predictions from chemical evolution models that include dust formation, growth, destruction, and dilution processes \citep[e.g.][]{feldmann2015}. This result is also consistent with the D/G---Z relation observed in nearby galaxies using the FIR to trace dust, 21 cm emission to trace atomic gas, and CO rotational emission to measure the mass of molecular gas \citep{remy-ruyer2014, devis2019}. However, the depletion-based D/G is a factor of a few (2 for the LMC, 5 for the SMC) higher than the D/G derived from FIR, 21 cm, and CO rotational emission. A combination of the uncertain dust FIR opacity, geometric effects, and the uncertainty of the depletion-based D/G due to the lack of C and O depletion measurements outside the MW, could explain this discrepancy. Nevertheless, both measurements are in agreement with chemical evolution models, owing to the metallicities of the LMC and SMC being higher than or on the cusp of the critical metallicity at which the D/M and D/G sharply drop. Below this critical metallicity of 10-20\% solar, the D/G predicted by chemical evolution models and observed in the FIR drops abruptly with metallicity. The models would interpret this finding as the growth of dust in the ISM becoming too inefficient below a critical metallicity of 10---20\% solar to counter-act the effects of dust destruction in supernova shocks and dust dilution from pristine inflows. No depletion measurements are yet available at those metallicities, but will soon be available from the METAL-Z large HST program (GO-15880, Hamanowicz et al., in prep).\\

\begin{acknowledgments}
We thank the referee for providing insightful comments and suggestions. Edward B. Jenkins, Benjamin Williams, Karl Gordon, Karin Sandstrom, and Petia Yanchulova Merica-Jones acknowledge support from grant HST-GO-14675. This work is based on observations with the NASA/ESA Hubble Space Telescope obtained at the Space Telescope Science Institute, which is operated by the Associations of Universities for Research in Astronomy, Incorporated, under NASA contract NAS5-26555. These observations are associated with program 14675. Support for Program number 14675 was provided by NASA through a grant from the Space Telescope Science Institute, which is operated by the Association of Universities for Research in Astronomy, Incorporated, under NASA contract NAS5-26555.
\end{acknowledgments}

\bibliography{/Users/duval/stsci_research/biblio_all}{}
\bibliographystyle{aasjournal}

\end{document}